\documentclass[preprint,prd,tightenlines,superscriptaddress,11pt]{revtex4}

\usepackage{graphicx} 
\usepackage{dcolumn}  
\usepackage{colordvi}
\usepackage{color}
\usepackage{epstopdf}
\usepackage{amssymb}
\usepackage{url}
\usepackage{rotating}

\usepackage{tabularx}
\usepackage{microtype}
\usepackage{hyperref}
\usepackage[italic]{hepnames}
\usepackage{hepparticles}
\usepackage{physics}
\usepackage{siunitx}
\usepackage{bm}
\usepackage[noabbrev, capitalise]{cleveref}

\usepackage[section]{placeins}

\graphicspath{figures/}

\newcommand{\PX}{\HepParticle{X}{}{}\xspace}
\newcommand{\PXu}{\HepParticle{X}{u}{}\xspace}
\newcommand{\PXc}{\HepParticle{X}{c}{}\xspace}
\newcommand{\PXs}{\HepParticle{X}{s}{}\xspace}

\newcommand{\PDstst}{\HepParticle{D}{}{**}\xspace}
\newcommand{\PDbostbc}{\HepParticle{D}{}{(*)}\xspace}

\newcommand{\PDone}{\HepParticle{D}{1}{}\xspace}
\newcommand{\PDtwost}{\HepParticle{D}{2}{\ast}\xspace}
\newcommand{\PDprimeone}{\HepParticle{D}{1}{\prime}\xspace}
\newcommand{\PDzerost}{\HepParticle{D}{0}{\ast}\xspace}

\newcommand{\BBpair}{\ensuremath{\PB\APB}\xspace}
\newcommand{\eepair}{\ensuremath{\APelectron\Pelectron}\xspace}

\newcommand{\Btag}{\HepParticle{\PB}{\mathrm{tag}}{}\xspace}

\newcommand{\elleqelmu}{\ensuremath{\Plepton = \Pe, \Pmu}\xspace}

\newcommand{\BtoXlv}{\HepProcess{\PB \to\PX \Plepton \Pgnl}\xspace}
\newcommand{\BtoXulv}{\HepProcess{\PB \to \PXu \Plepton \Pgnl}\xspace}
\newcommand{\BtoXclv}{\HepProcess{\PB \to \PXc \Plepton \Pgnl}\xspace}

\newcommand{\BtoDlnu}{\HepProcess{\PB\to\PD\Plepton\Pgnl}\xspace}

\newcommand{\BtoDstlnu}{\HepProcess{\PB\to\PDst\Plepton\Pgnl}\xspace}

\newcommand{\YFourStoBB}{\HepProcess{\PUpsilonFourS \to \PB \APB}\xspace}

\newcommand{\eetocont}{\HepProcess{\eepair\to\Pquark \APquark} ($\Pquark =  \Pqu,\Pqd,\Pqs,\Pqc$)\xspace}

\newcommand{\mx}{\ensuremath{M_{\PX}}\xspace}
\newcommand{\mnx}[1]{\ensuremath{M^{#1}_{\PX}}\xspace}

\newcommand{\mnxcalib}[1]{\ensuremath{M^{#1}_{\PX,\mathrm{calib}}}\xspace}

\newcommand{\mnxcalibevent}[1]{\ensuremath{M^{#1}_{\PX,\mathrm{calib},i}}\xspace}

\newcommand{\mxreco}{\ensuremath{M_{\PX,\mathrm{reco}}}\xspace}
\newcommand{\mxcalib}{\ensuremath{M_{\PX,\mathrm{calib}}}\xspace}
\newcommand{\mxtrue}{\ensuremath{M_{\PX,\mathrm{true}}}\xspace}

\newcommand{\mxntruesignal}{\ensuremath{M^n_{\PX,\mathrm{true,signal}}}\xspace}

\newcommand{\mxmoments}{\ensuremath{\langle \mx^{n} \rangle}\xspace}

\newcommand{\mxmomentscalib}{\ensuremath{\langle \mxcalib^n \rangle}\xspace}
\newcommand{\mxmomentstrue}{\ensuremath{\langle \mxtrue^n \rangle}\xspace}
\newcommand{\mxmomentstruesignal}{\ensuremath{\langle \mxntruesignal\rangle}\xspace}

\newcommand{\mxmoment}[1]{\ensuremath{\langle \mx^{#1} \rangle}\xspace}
\newcommand{\mxmomentreco}[1]{\ensuremath{\langle \mxreco^{#1}  \rangle}\xspace}

\newcommand{\mxmomenttrue}[1]{\ensuremath{\langle \mxtrue^{#1}  \rangle}\xspace}

\newcommand{\ccalib}{\ensuremath{\mathcal{C}_\mathrm{calib}}\xspace}
\newcommand{\ctrue}{\ensuremath{\mathcal{C}_\mathrm{true}}\xspace}

\newcommand{\Xmultiplicity}{\ensuremath{\mathrm{X_{mult}}}\xspace}

\newcommand{\plepsigbrestframe}{\ensuremath{p^\ast_\ell}\xspace}
\newcommand{\pmiss}{\ensuremath{p_\mathrm{miss}}\xspace}
\newcommand{\cpmiss}{\ensuremath{c\cdot p_\mathrm{miss}}\xspace}
\newcommand{\Emiss}{\ensuremath{E_\mathrm{miss}}\xspace}
\newcommand{\Emissminuspmiss}{\ensuremath{ \Emiss -\pmiss} \xspace}
\newcommand{\Emissminuscpmiss}{\ensuremath{ \Emiss -\cpmiss} \xspace}

\newcommand{\absEmissminuscpmiss}{\ensuremath{\left| \Emissminuscpmiss \right|}\xspace}
\newcommand{\totalCharge}{\ensuremath{Q_\mathrm{tot}}\xspace}
\newcommand{\abstotalCharge}{\ensuremath{\left| \totalCharge \right|}\xspace}

\newcommand{\mbc}{\ensuremath{M_{\mathrm{bc}}}\xspace}

\newcommand{\DeltaE}{\ensuremath{\Delta E}\xspace}

\newcommand{\FEIprob}{\ensuremath{\mathcal{P}_\mathrm{FEI}}\xspace}
\newcommand{\CSprob}{\ensuremath{\mathcal{P}_\mathrm{CS}}\xspace}

\newcommand{\dr}{\ensuremath{dr}\xspace}
\newcommand{\dz}{\ensuremath{\left| dz \right|}\xspace}

\newcommand{\pt}{\ensuremath{p_T}\xspace}

\newcommand{\Rtwo}{\ensuremath{R_2}\xspace}

\newcommand{\Belle}{Belle\xspace}

\newcommand{\FEI}{FEI\xspace}

\begin{document}

\vspace*{-3\baselineskip}
\resizebox{!}{3cm}{\includegraphics{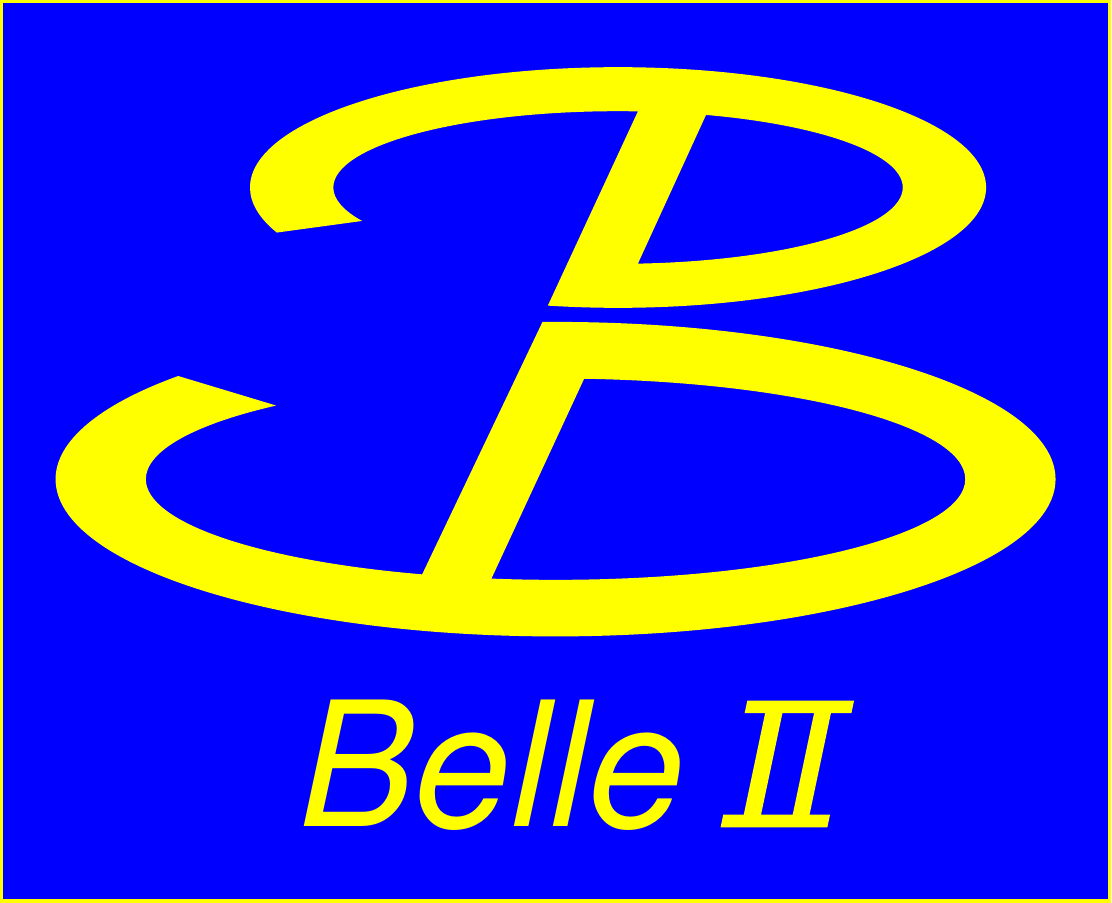}}

\vspace*{-3\baselineskip}
\begin{flushright}
BELLE2-CONF-PH-2020-011\\
September 9, 2020
\end{flushright}

\title {\vspace{0.5cm} \quad\\[0.5cm] Measurement of Hadronic Mass Moments \mxmoments in \BtoXclv Decays at Belle~II}

\newcommand{\instSinica}{Academia Sinica, Taipei 11529, Taiwan}
\newcommand{\instCPPM}{Aix Marseille Universit\'{e}, CNRS/IN2P3, CPPM, 13288 Marseille, France}
\newcommand{\instBeihang}{Beihang University, Beijing 100191, China}
\newcommand{\instBUAP}{Benemerita Universidad Autonoma de Puebla, Puebla 72570, Mexico}
\newcommand{\instBNL}{Brookhaven National Laboratory, Upton, New York 11973, U.S.A.}
\newcommand{\instBINP}{Budker Institute of Nuclear Physics SB RAS, Novosibirsk 630090, Russian Federation}
\newcommand{\instCMU}{Carnegie Mellon University, Pittsburgh, Pennsylvania 15213, U.S.A.}
\newcommand{\instCinvestavIPN}{Centro de Investigacion y de Estudios Avanzados del Instituto Politecnico Nacional, Mexico City 07360, Mexico}
\newcommand{\instPrague}{Faculty of Mathematics and Physics, Charles University, 121 16 Prague, Czech Republic}
\newcommand{\instChiangMai}{Chiang Mai University, Chiang Mai 50202, Thailand}
\newcommand{\instChiba}{Chiba University, Chiba 263-8522, Japan}
\newcommand{\instChonnam}{Chonnam National University, Gwangju 61186, South Korea}
\newcommand{\instConacyt}{Consejo Nacional de Ciencia y Tecnolog\'{\i}a, Mexico City 03940, Mexico}
\newcommand{\instDESY}{Deutsches Elektronen--Synchrotron, 22607 Hamburg, Germany}
\newcommand{\instDuke}{Duke University, Durham, North Carolina 27708, U.S.A.}
\newcommand{\instITAR}{Institute of Theoretical and Applied Research (ITAR), Duy Tan University, Hanoi 100000, Vietnam}
\newcommand{\instENEA}{ENEA Casaccia, I-00123 Roma, Italy}
\newcommand{\instEri}{Earthquake Research Institute, University of Tokyo, Tokyo 113-0032, Japan}
\newcommand{\instJuelich}{Forschungszentrum J\"{u}lich, 52425 J\"{u}lich, Germany}
\newcommand{\instFuJen}{Department of Physics, Fu Jen Catholic University, Taipei 24205, Taiwan}
\newcommand{\instFudan}{Key Laboratory of Nuclear Physics and Ion-beam Application (MOE) and Institute of Modern Physics, Fudan University, Shanghai 200443, China}
\newcommand{\instGoettingen}{II. Physikalisches Institut, Georg-August-Universit\"{a}t G\"{o}ttingen, 37073 G\"{o}ttingen, Germany}
\newcommand{\instGifu}{Gifu University, Gifu 501-1193, Japan}
\newcommand{\instSOKENDAI}{The Graduate University for Advanced Studies (SOKENDAI), Hayama 240-0193, Japan}
\newcommand{\instGyeongsang}{Gyeongsang National University, Jinju 52828, South Korea}
\newcommand{\instHanyang}{Department of Physics and Institute of Natural Sciences, Hanyang University, Seoul 04763, South Korea}
\newcommand{\instKEK}{High Energy Accelerator Research Organization (KEK), Tsukuba 305-0801, Japan}
\newcommand{\instJPARC}{J-PARC Branch, KEK Theory Center, High Energy Accelerator Research Organization (KEK), Tsukuba 305-0801, Japan}
\newcommand{\instHSE}{Higher School of Economics (HSE), Moscow 101000, Russian Federation}
\newcommand{\instIISER}{Indian Institute of Science Education and Research Mohali, SAS Nagar, 140306, India}
\newcommand{\instIITBhubaneswar}{Indian Institute of Technology Bhubaneswar, Satya Nagar 751007, India}
\newcommand{\instIITGuwahati}{Indian Institute of Technology Guwahati, Assam 781039, India}
\newcommand{\instIITHyderabad}{Indian Institute of Technology Hyderabad, Telangana 502285, India}
\newcommand{\instIITMadras}{Indian Institute of Technology Madras, Chennai 600036, India}
\newcommand{\instIndiana}{Indiana University, Bloomington, Indiana 47408, U.S.A.}
\newcommand{\instIHEPRussia}{Institute for High Energy Physics, Protvino 142281, Russian Federation}
\newcommand{\instHEPHYVienna}{Institute of High Energy Physics, Vienna 1050, Austria}
\newcommand{\instIHEPChina}{Institute of High Energy Physics, Chinese Academy of Sciences, Beijing 100049, China}
\newcommand{\instChennai}{Institute of Mathematical Sciences, Chennai 600113, India}
\newcommand{\instIPP}{Institute of Particle Physics (Canada), Victoria, British Columbia V8W 2Y2, Canada}
\newcommand{\instIOP}{Institute of Physics, Vietnam Academy of Science and Technology (VAST), Hanoi, Vietnam}
\newcommand{\instIFIC}{Instituto de Fisica Corpuscular, Paterna 46980, Spain}
\newcommand{\instFrascati}{INFN Laboratori Nazionali di Frascati, I-00044 Frascati, Italy}
\newcommand{\instNapoliINFN}{INFN Sezione di Napoli, I-80126 Napoli, Italy}
\newcommand{\instPadovaINFN}{INFN Sezione di Padova, I-35131 Padova, Italy}
\newcommand{\instPerugiaINFN}{INFN Sezione di Perugia, I-06123 Perugia, Italy}
\newcommand{\instPisaINFN}{INFN Sezione di Pisa, I-56127 Pisa, Italy}
\newcommand{\instRomaINFN}{INFN Sezione di Roma, I-00185 Roma, Italy}
\newcommand{\instRomaTreINFN}{INFN Sezione di Roma Tre, I-00146 Roma, Italy}
\newcommand{\instTorinoINFN}{INFN Sezione di Torino, I-10125 Torino, Italy}
\newcommand{\instTriesteINFN}{INFN Sezione di Trieste, I-34127 Trieste, Italy}
\newcommand{\instJAEA}{Advanced Science Research Center, Japan Atomic Energy Agency, Naka 319-1195, Japan}
\newcommand{\instMainz}{Johannes Gutenberg-Universit\"{a}t Mainz, Institut f\"{u}r Kernphysik, D-55099 Mainz, Germany}
\newcommand{\instGiessen}{Justus-Liebig-Universit\"{a}t Gie\ss{}en, 35392 Gie\ss{}en, Germany}
\newcommand{\instKarlsruhe}{Institut f\"{u}r Experimentelle Teilchenphysik, Karlsruher Institut f\"{u}r Technologie, 76131 Karlsruhe, Germany}
\newcommand{\instKennesaw}{Kennesaw State University, Kennesaw, Georgia 30144, U.S.A.}
\newcommand{\instKitasato}{Kitasato University, Sagamihara 252-0373, Japan}
\newcommand{\instKISTI}{Korea Institute of Science and Technology Information, Daejeon 34141, South Korea}
\newcommand{\instKorea}{Korea University, Seoul 02841, South Korea}
\newcommand{\instKSU}{Kyoto Sangyo University, Kyoto 603-8555, Japan}
\newcommand{\instKyotoU}{Kyoto University, Kyoto 606-8501, Japan}
\newcommand{\instKyungpook}{Kyungpook National University, Daegu 41566, South Korea}
\newcommand{\instLPI}{P.N. Lebedev Physical Institute of the Russian Academy of Sciences, Moscow 119991, Russian Federation}
\newcommand{\instLNNU}{Liaoning Normal University, Dalian 116029, China}
\newcommand{\instLMU}{Ludwig Maximilians University, 80539 Munich, Germany}
\newcommand{\instLuther}{Luther College, Decorah, Iowa 52101, U.S.A.}
\newcommand{\instMNITJaipur}{Malaviya National Institute of Technology Jaipur, Jaipur 302017, India}
\newcommand{\instMPP}{Max-Planck-Institut f\"{u}r Physik, 80805 M\"{u}nchen, Germany}
\newcommand{\instMPGHLL}{Semiconductor Laboratory of the Max Planck Society, 81739 M\"{u}nchen, Germany}
\newcommand{\instMcGill}{McGill University, Montr\'{e}al, Qu\'{e}bec, H3A 2T8, Canada}
\newcommand{\instMETU}{Middle East Technical University, 06531 Ankara, Turkey}
\newcommand{\instMEPhI}{Moscow Physical Engineering Institute, Moscow 115409, Russian Federation}
\newcommand{\instNagoya}{Graduate School of Science, Nagoya University, Nagoya 464-8602, Japan}
\newcommand{\instNagoyaKMI}{Kobayashi-Maskawa Institute, Nagoya University, Nagoya 464-8602, Japan}
\newcommand{\instNagoyaIAR}{Institute for Advanced Research, Nagoya University, Nagoya 464-8602, Japan}
\newcommand{\instNaraWu}{Nara Women's University, Nara 630-8506, Japan}
\newcommand{\instUNAM}{National Autonomous University of Mexico, Mexico City, Mexico}
\newcommand{\instNTUTaiwan}{Department of Physics, National Taiwan University, Taipei 10617, Taiwan}
\newcommand{\instNUUTaiwan}{National United University, Miao Li 36003, Taiwan}
\newcommand{\instKrakow}{H. Niewodniczanski Institute of Nuclear Physics, Krakow 31-342, Poland}
\newcommand{\instNiigata}{Niigata University, Niigata 950-2181, Japan}
\newcommand{\instNSU}{Novosibirsk State University, Novosibirsk 630090, Russian Federation}
\newcommand{\instOkinawa}{Okinawa Institute of Science and Technology, Okinawa 904-0495, Japan}
\newcommand{\instOsakaCity}{Osaka City University, Osaka 558-8585, Japan}
\newcommand{\instRCNP}{Research Center for Nuclear Physics, Osaka University, Osaka 567-0047, Japan}
\newcommand{\instPNNL}{Pacific Northwest National Laboratory, Richland, Washington 99352, U.S.A.}
\newcommand{\instPanjab}{Panjab University, Chandigarh 160014, India}
\newcommand{\instPeking}{Peking University, Beijing 100871, China}
\newcommand{\instPanjabPAU}{Punjab Agricultural University, Ludhiana 141004, India}
\newcommand{\instRIKENMSL}{Meson Science Laboratory, Cluster for Pioneering Research, RIKEN, Saitama 351-0198, Japan}
\newcommand{\instRIKEN}{Theoretical Research Division, Nishina Center, RIKEN, Saitama 351-0198, Japan}
\newcommand{\instXavier}{St. Francis Xavier University, Antigonish, Nova Scotia, B2G 2W5, Canada}
\newcommand{\instSeoul}{Seoul National University, Seoul 08826, South Korea}
\newcommand{\instShandong}{Shandong University, Jinan 250100, China}
\newcommand{\instSPU}{Showa Pharmaceutical University, Tokyo 194-8543, Japan}
\newcommand{\instSoochow}{Soochow University, Suzhou 215006, China}
\newcommand{\instSoongsil}{Soongsil University, Seoul 06978, South Korea}
\newcommand{\instLjubljanaJSI}{J. Stefan Institute, 1000 Ljubljana, Slovenia}
\newcommand{\instKyiv}{Taras Shevchenko National Univ. of Kiev, Kiev, Ukraine}
\newcommand{\instTata}{Tata Institute of Fundamental Research, Mumbai 400005, India}
\newcommand{\instTUM}{Department of Physics, Technische Universit\"{a}t M\"{u}nchen, 85748 Garching, Germany}
\newcommand{\instECUTUM}{Excellence Cluster Universe, Technische Universit\"{a}t M\"{u}nchen, 85748 Garching, Germany}
\newcommand{\instTelAviv}{Tel Aviv University, School of Physics and Astronomy, Tel Aviv, 69978, Israel}
\newcommand{\instToho}{Toho University, Funabashi 274-8510, Japan}
\newcommand{\instTohoku}{Department of Physics, Tohoku University, Sendai 980-8578, Japan}
\newcommand{\instTitech}{Tokyo Institute of Technology, Tokyo 152-8550, Japan}
\newcommand{\instTokyoMetropolitan}{Tokyo Metropolitan University, Tokyo 192-0397, Japan}
\newcommand{\instUAS}{Universidad Autonoma de Sinaloa, Sinaloa 80000, Mexico}
\newcommand{\instNapoliUNIV}{Dipartimento di Scienze Fisiche, Universit\`{a} di Napoli Federico II, I-80126 Napoli, Italy}
\newcommand{\instNapoliUNIVA}{Dipartimento di Agraria, Universit\`{a} di Napoli Federico II, I-80055 Portici (NA), Italy}
\newcommand{\instPadovaUNIV}{Dipartimento di Fisica e Astronomia, Universit\`{a} di Padova, I-35131 Padova, Italy}
\newcommand{\instPerugiaUNIV}{Dipartimento di Fisica, Universit\`{a} di Perugia, I-06123 Perugia, Italy}
\newcommand{\instPisaUNIV}{Dipartimento di Fisica, Universit\`{a} di Pisa, I-56127 Pisa, Italy}
\newcommand{\instRomaUNIV}{Universit\`{a} di Roma ``La Sapienza,'' I-00185 Roma, Italy}
\newcommand{\instRomaTreUNIV}{Dipartimento di Matematica e Fisica, Universit\`{a} di Roma Tre, I-00146 Roma, Italy}
\newcommand{\instTorinoUNIV}{Dipartimento di Fisica, Universit\`{a} di Torino, I-10125 Torino, Italy}
\newcommand{\instTriesteUNIV}{Dipartimento di Fisica, Universit\`{a} di Trieste, I-34127 Trieste, Italy}
\newcommand{\instMontreal}{Universit\'{e} de Montr\'{e}al, Physique des Particules, Montr\'{e}al, Qu\'{e}bec, H3C 3J7, Canada}
\newcommand{\instIJCLab}{Universit\'{e} Paris-Saclay, CNRS/IN2P3, IJCLab, 91405 Orsay, France}
\newcommand{\instIPHC}{Universit\'{e} de Strasbourg, CNRS, IPHC, UMR 7178, 67037 Strasbourg, France}
\newcommand{\instAdelaide}{Department of Physics, University of Adelaide, Adelaide, South Australia 5005, Australia}
\newcommand{\instBonn}{University of Bonn, 53115 Bonn, Germany}
\newcommand{\instUBC}{University of British Columbia, Vancouver, British Columbia, V6T 1Z1, Canada}
\newcommand{\instCincinnati}{University of Cincinnati, Cincinnati, Ohio 45221, U.S.A.}
\newcommand{\instFlorida}{University of Florida, Gainesville, Florida 32611, U.S.A.}
\newcommand{\instHamburg}{University of Hamburg, 20148 Hamburg, Germany}
\newcommand{\instHawaii}{University of Hawaii, Honolulu, Hawaii 96822, U.S.A.}
\newcommand{\instHeidelberg}{University of Heidelberg, 68131 Mannheim, Germany}
\newcommand{\instLjubljanaUniLJ}{Faculty of Mathematics and Physics, University of Ljubljana, 1000 Ljubljana, Slovenia}
\newcommand{\instLouisville}{University of Louisville, Louisville, Kentucky 40292, U.S.A.}
\newcommand{\instMalaya}{National Centre for Particle Physics, University Malaya, 50603 Kuala Lumpur, Malaysia}
\newcommand{\instLjubljanaUM}{University of Maribor, 2000 Maribor, Slovenia}
\newcommand{\instMelbourne}{School of Physics, University of Melbourne, Victoria 3010, Australia}
\newcommand{\instMississippi}{University of Mississippi, University, Mississippi 38677, U.S.A.}
\newcommand{\instUOM}{University of Miyazaki, Miyazaki 889-2192, Japan}
\newcommand{\instNovaGorica}{University of Nova Gorica, 5000 Nova Gorica, Slovenia}
\newcommand{\instPittsburgh}{University of Pittsburgh, Pittsburgh, Pennsylvania 15260, U.S.A.}
\newcommand{\instUSTC}{University of Science and Technology of China, Hefei 230026, China}
\newcommand{\instSAlabama}{University of South Alabama, Mobile, Alabama 36688, U.S.A.}
\newcommand{\instSCarolina}{University of South Carolina, Columbia, South Carolina 29208, U.S.A.}
\newcommand{\instSydney}{School of Physics, University of Sydney, New South Wales 2006, Australia}
\newcommand{\instTabuk}{Department of Physics, Faculty of Science, University of Tabuk, Tabuk 71451, Saudi Arabia}
\newcommand{\instUTokyo}{Department of Physics, University of Tokyo, Tokyo 113-0033, Japan}
\newcommand{\instIPMU}{Kavli Institute for the Physics and Mathematics of the Universe (WPI), University of Tokyo, Kashiwa 277-8583, Japan}
\newcommand{\instVictoria}{University of Victoria, Victoria, British Columbia, V8W 3P6, Canada}
\newcommand{\instVPI}{Virginia Polytechnic Institute and State University, Blacksburg, Virginia 24061, U.S.A.}
\newcommand{\instWayneState}{Wayne State University, Detroit, Michigan 48202, U.S.A.}
\newcommand{\instYamagata}{Yamagata University, Yamagata 990-8560, Japan}
\newcommand{\instYerevan}{Alikhanyan National Science Laboratory, Yerevan 0036, Armenia}
\newcommand{\instYonsei}{Yonsei University, Seoul 03722, South Korea}
\affiliation{\instCPPM}
\affiliation{\instBeihang}
\affiliation{\instBNL}
\affiliation{\instBINP}
\affiliation{\instCMU}
\affiliation{\instCinvestavIPN}
\affiliation{\instPrague}
\affiliation{\instChiangMai}
\affiliation{\instChiba}
\affiliation{\instChonnam}
\affiliation{\instConacyt}
\affiliation{\instDESY}
\affiliation{\instDuke}
\affiliation{\instITAR}
\affiliation{\instEri}
\affiliation{\instJuelich}
\affiliation{\instFuJen}
\affiliation{\instFudan}
\affiliation{\instGoettingen}
\affiliation{\instGifu}
\affiliation{\instSOKENDAI}
\affiliation{\instGyeongsang}
\affiliation{\instHanyang}
\affiliation{\instKEK}
\affiliation{\instJPARC}
\affiliation{\instHSE}
\affiliation{\instIISER}
\affiliation{\instIITBhubaneswar}
\affiliation{\instIITGuwahati}
\affiliation{\instIITHyderabad}
\affiliation{\instIITMadras}
\affiliation{\instIndiana}
\affiliation{\instIHEPRussia}
\affiliation{\instHEPHYVienna}
\affiliation{\instIHEPChina}
\affiliation{\instIPP}
\affiliation{\instIOP}
\affiliation{\instIFIC}
\affiliation{\instFrascati}
\affiliation{\instNapoliINFN}
\affiliation{\instPadovaINFN}
\affiliation{\instPerugiaINFN}
\affiliation{\instPisaINFN}
\affiliation{\instRomaINFN}
\affiliation{\instRomaTreINFN}
\affiliation{\instTorinoINFN}
\affiliation{\instTriesteINFN}
\affiliation{\instJAEA}
\affiliation{\instMainz}
\affiliation{\instGiessen}
\affiliation{\instKarlsruhe}
\affiliation{\instKitasato}
\affiliation{\instKISTI}
\affiliation{\instKorea}
\affiliation{\instKSU}
\affiliation{\instKyungpook}
\affiliation{\instLPI}
\affiliation{\instLNNU}
\affiliation{\instLMU}
\affiliation{\instLuther}
\affiliation{\instMNITJaipur}
\affiliation{\instMPP}
\affiliation{\instMPGHLL}
\affiliation{\instMcGill}
\affiliation{\instMEPhI}
\affiliation{\instNagoya}
\affiliation{\instNagoyaKMI}
\affiliation{\instNagoyaIAR}
\affiliation{\instNaraWu}
\affiliation{\instNTUTaiwan}
\affiliation{\instNUUTaiwan}
\affiliation{\instKrakow}
\affiliation{\instNiigata}
\affiliation{\instNSU}
\affiliation{\instOkinawa}
\affiliation{\instOsakaCity}
\affiliation{\instRCNP}
\affiliation{\instPNNL}
\affiliation{\instPanjab}
\affiliation{\instPeking}
\affiliation{\instPanjabPAU}
\affiliation{\instRIKENMSL}
\affiliation{\instSeoul}
\affiliation{\instSPU}
\affiliation{\instSoochow}
\affiliation{\instSoongsil}
\affiliation{\instLjubljanaJSI}
\affiliation{\instKyiv}
\affiliation{\instTata}
\affiliation{\instTUM}
\affiliation{\instTelAviv}
\affiliation{\instToho}
\affiliation{\instTohoku}
\affiliation{\instTitech}
\affiliation{\instTokyoMetropolitan}
\affiliation{\instUAS}
\affiliation{\instNapoliUNIV}
\affiliation{\instPadovaUNIV}
\affiliation{\instPerugiaUNIV}
\affiliation{\instPisaUNIV}
\affiliation{\instRomaUNIV}
\affiliation{\instRomaTreUNIV}
\affiliation{\instTorinoUNIV}
\affiliation{\instTriesteUNIV}
\affiliation{\instMontreal}
\affiliation{\instIJCLab}
\affiliation{\instIPHC}
\affiliation{\instAdelaide}
\affiliation{\instBonn}
\affiliation{\instUBC}
\affiliation{\instCincinnati}
\affiliation{\instFlorida}
\affiliation{\instHawaii}
\affiliation{\instHeidelberg}
\affiliation{\instLjubljanaUniLJ}
\affiliation{\instLouisville}
\affiliation{\instMalaya}
\affiliation{\instLjubljanaUM}
\affiliation{\instMelbourne}
\affiliation{\instMississippi}
\affiliation{\instUOM}
\affiliation{\instPittsburgh}
\affiliation{\instUSTC}
\affiliation{\instSAlabama}
\affiliation{\instSCarolina}
\affiliation{\instSydney}
\affiliation{\instUTokyo}
\affiliation{\instIPMU}
\affiliation{\instVictoria}
\affiliation{\instVPI}
\affiliation{\instWayneState}
\affiliation{\instYamagata}
\affiliation{\instYerevan}
\affiliation{\instYonsei}
  \author{F.~Abudin{\'e}n}\affiliation{\instTriesteINFN} 
  \author{I.~Adachi}\affiliation{\instKEK}\affiliation{\instSOKENDAI} 
  \author{R.~Adak}\affiliation{\instFudan} 
  \author{K.~Adamczyk}\affiliation{\instKrakow} 
  \author{P.~Ahlburg}\affiliation{\instBonn} 
  \author{J.~K.~Ahn}\affiliation{\instKorea} 
  \author{H.~Aihara}\affiliation{\instUTokyo} 
  \author{N.~Akopov}\affiliation{\instYerevan} 
  \author{A.~Aloisio}\affiliation{\instNapoliUNIV}\affiliation{\instNapoliINFN} 
  \author{F.~Ameli}\affiliation{\instRomaINFN} 
% ...
\collaboration{Belle II Collaboration}

\begin{abstract}
    
    We present measurements of the first six hadronic mass moments in semileptonic \BtoXclv decays.
    The hadronic mass moments, together with other observables of inclusive \PB decays, can be used to determine the CKM matrix element $\abs{V_{\Pqc \Pqb}}$ and mass of the $\Pqb$-quark $m_{\Pqb}$ in the context of Heavy Quark Expansions of QCD.
    The Belle~II data recorded at the \PUpsilonFourS resonance in 2019 and 2020 (March-July), corresponding to an integrated luminosity of \SI{34.6}{fb^{-1}}, is used for this measurement.
    The decay \YFourStoBB is reconstructed by applying the hadronic tagging algorithm provided by the Full Event Interpretation to fully reconstruct one \PB meson. The second \PB meson is reconstructed inclusively by selecting a high-momentum lepton.
    The $X_c$ system is identified by the remaining reconstructed tracks and clusters in the electromagnetic calorimeter.
    We report preliminary results for the hadronic mass moments \mxmoments with $n=1,\dots,6$, measured as a function of a lower cut on the lepton momentum in the signal \PB rest frame.

\end{abstract}


\maketitle
\normalsize 
{\renewcommand{\thefootnote}{\fnsymbol{footnote}}}
\setcounter{footnote}{0}
\section{Introduction}
The mass moments \mxmoments of the hadronic system in inclusive semileptonic \BtoXclv decays  can be used to measure non-perturbative QCD parameters and the CKM matrix element $\abs{V_{cb}}$. The state-of-the-art procedure relies on combining the information from mass moments, with measured moments from the lepton energy spectrum and \HepProcess{\PB \to \PXs \Pgamma} information, to perform a combined fit using theory predictions building on the Heavy Quark Expansions of QCD to determine $\abs{V_{cb}}$ and the $b$ quark mass $m_{\Pqb}$.
See e.g. Ref.~\cite{Gambino:2020jvv} for a recent review.

This work presents the first results of hadronic mass moments \mxmoments with $n=1,\dots,6$, measured at the \Belle~II experiment.
In this analysis, semileptonic \BtoXclv decays are reconstructed inclusively by selecting a high-momentum lepton.
The other \PB meson is fully reconstructed in hadronic modes via the Full Event Interpretation (\FEI) \cite{Keck_2019}.
This \PB meson is referred to as the tag-side \PB meson (\Btag) throughout this note.
We subtract the remaining background components by assigning a continuous signal probability as a function of the reconstructed mass of the hadronic \PXc system (\mx) to each event.
A calibration procedure is applied to correct for a bias in the reconstructed \mx spectrum due to experimental effects.
The hadronic mass moments are calculated as a weighted mean of the calibrated \mx distribution, where the events are weighted with the aforementioned signal probability.

The rest of this note is organized as follows.
\cref{sec:detector} briefly describes the Belle~II detector and how the inclusive \BtoXclv decays are simulated.
The reconstruction of the \PUpsilonFourS events is discussed in \cref{sec:reconstruction}.
The procedure for subtracting remaining background components from the measured \mx spectrum is introduced in \cref{sec:bkg_subtraction}.
\cref{sec:mx_moments_measurements} discusses the extraction and calibration of the reconstructed \mx distributions. In addition, the handling of statistical and systematic uncertainties is explained and the measured \mx values are given. Finally, \cref{sec:summary} presents our conclusions.

\section{Belle II Detector and Data Set}
\label{sec:detector}
The Belle~II  detector \cite{Abe:2010sj} is operated at the SuperKEKB electron-positron collider \cite{Akai:2018mbz} and is located at the KEK laboratory in Tsukuba, Japan. 
The  detector consists of several nested detector subsystems arranged around the beam pipe in a cylindrical geometry.
Sub-detectors relevant for this analysis are briefly described here; a description of the full detector is given in \cite{Abe:2010sj,Kou:2018nap}.
The innermost subsystem is the vertex detector, which includes two layers of silicon pixel detectors and four outer layers of silicon strip detectors.
Currently, the second pixel layer is installed to cover only a small part of the solid angle, while the remaining vertex detector layers are fully installed.
Most of the tracking volume consists of a helium and ethane-based small-cell drift chamber.
Surrounding the drift chamber (CDC), the Cherenkov-light imaging and time-of-propagation detector provides charged-particle identification in the barrel region. In the forward end-cap, this function is provided by a proximity-focusing, ring-imaging Cherenkov detector with an aerogel radiator. 
The next sub-detector layer consists of the electromagnetic calorimeter (ECL), composed of barrel and two end-cap sections made of CsI(Tl) crystals.
The inner detector is immersed in a uniform magnetic field with a field strength of \SI{1.5}{T} from the superconducting solenoid situated outside the calorimeter.
Multiple layers of scintillators and resistive plate chambers, located between the magnetic flux-return iron plates, constitute the \PKlong and muon identification system.


The data sample used in this analysis was collected in 2019 and from March to July 2020 at a center-of-mass (CM) energy of $\sqrt{s} = \SI{10.58}{GeV}$, corresponding to the mass of the \PUpsilonFourS resonance. 
The energies of the electron and positron beams are \SI{7}{GeV} and \SI{4}{GeV}, respectively, resulting in a boost of $\beta\gamma=0.28$ of the CM frame relative to the laboratory frame. 
The integrated luminosity of the data sample amounts to \SI{34.6}{fb^{-1}}.

Monte Carlo (MC) samples of \PB meson decays are simulated using the \texttt{EvtGen} generator \cite{LANGE2001152}. 
The sample size corresponds to an integrated luminosity of \SI{200}{fb^{-1}}.
The interactions of particles inside the detector are simulated using \texttt{Geant4} \cite{Agostinelli:2002hh}.
Electromagnetic final-state radiation (FSR) is simulated using the \texttt{PHOTOS} \cite{BARBERIO1991115} package.
The simulation of the continuum background process \eetocont is carried out with \texttt{KKMC}~\cite{Ward:2002qq}, interfaced with \texttt{Pythia}~\cite{Sj_strand_2008}.
All recorded collisions and simulated events were analyzed in the \texttt{basf2} framework~\cite{Kuhr:2018lps} and a summary of the track and ECL reconstruction algorithms can be found in Ref.~\cite{Bertacchi:2020eez} and Ref.~\cite{Kou:2018nap}, respectively.

The \BtoXclv spectrum is modeled as a mixture of resonant and non-resonant decays.
\BtoDlnu decays are modeled using the BGL form factors \cite{Boyd_1997} with central values taken from the fit in Ref. \cite{Glattauer_2016}.
To simulate \BtoDstlnu decays, the CLN form factors \cite{CAPRINI1998153} are used with central values taken from Ref. \cite{Amhis_2017}.
The decays of the four orbitally excited \PD meson states (\PDone, \PDtwost, \PDprimeone and \PDzerost), denoted as \PDstst, are simulated with a LLSW form factor inspired parametrization~\cite{Leibovich_1998}, using the central values and parametrization from Ref. \cite{Bernlochner_2017}. 
The non-resonant part of the \PXc spectrum is simulated as a composition of \HepProcess{\PB\to\PDbostbc \Ppi \Plepton\Pgnl}, \HepProcess{\PB \to \PDbostbc \Ppi  \Ppi  \Plepton \Pgnl} and \HepProcess{\PB \to \PDbostbc \Peta \Plepton \Pgnl} decays.
The first decay is simulated using the decay model proposed by Goity and Roberts \cite{Goity_1995}, while the remaining two decays are modeled with a pure phase-space prescription.
The branching fractions used for the simulation of \BtoXclv decays are given in \cref{tab:bfs}.

\begin{table}[tb]
\caption{
    Branching fractions used in the simulation of \BtoXclv decays in this analysis
}

\vspace{0.2cm}

\label{tab:bfs}
\begin{tabular}{lrr}
      \toprule
       $\mathcal{B}$ & Value $B^+$ & Value $B^0$ \\
       \colrule
       $B \to D \, \ell^+ \, \nu_\ell$ & $\left(2.3 \pm 0.1\right) \times 10^{-2} $ & $\left(2.1 \pm 0.1\right)\times 10^{-2} $ \\
       $B \to D^* \, \ell^+ \, \nu_\ell$ & $\left(5.5 \pm 0.1 \right)\times 10^{-2} $ &$\left( 5.1 \pm 0.1 \right)\times 10^{-2} $ \\
       
      \colrule
       $B \to D_1 \, \ell^+ \, \nu_\ell$ & $\left(4.5 \pm 0.3\right) \times 10^{-3}$ & $\left(4.2 \pm 0.3\right) \times 10^{-3}$ \\
       \quad\, $(\hookrightarrow D^* \pi)$ \\
       $B \to D_1 \, \ell^+ \, \nu_\ell$ & $\left(3.2 \pm 1.0\right) \times 10^{-3}$ & $\left(2.8 \pm 0.9\right) \times 10^{-3}$ \\
       \quad\, $(\hookrightarrow D \pi \pi)$ \\
        $B \to D_2^* \, \ell^+ \, \nu_\ell$ & $\left(1.5 \pm 0.1\right) \times 10^{-3}$ & $\left(1.4 \pm 0.1\right) \times 10^{-3}$  \\
       \quad\, $(\hookrightarrow D^* \pi)$ \\ 
        $B \to D_2^* \, \ell^+ \, \nu_\ell$ & $\left(2.2 \pm 0.2\right) \times 10^{-3}$ & $\left(2.1 \pm 0.2\right) \times 10^{-3}$  \\
       \quad\, $(\hookrightarrow D \pi)$ \\ 

       $B \to D_0^* \, \ell^+ \, \nu_\ell$ & $\left(3.9 \pm 0.8\right) \times 10^{-3}$ & $\left(3.6 \pm 0.7\right) \times 10^{-3}$ \\
       \quad\, $(\hookrightarrow D \pi)$ \\ 
        $B \to D_1^\prime \, \ell^+ \, \nu_\ell$ & $\left(4.3 \pm 0.8\right) \times 10^{-3}$ & $\left(4.0 \pm 0.8\right) \times 10^{-3}$  \\
        \quad\, $(\hookrightarrow D^* \pi)$ \\
      \colrule
       
        
       $B \to D \pi \, \ell^+ \, \nu_\ell$ & $\left(1.5 \pm 0.6 \right) \times 10^{-3}$ & $\left(1.5 \pm 0.6 \right) \times 10^{-3}$ \\
         $B \to D^* \pi \, \ell^+ \, \nu_\ell$ & $\left(1.5 \pm 1.0 \right) \times 10^{-3}$ & $\left(1.5 \pm 1.0 \right) \times 10^{-3}$ \\
        $B \to D \pi \pi \, \ell^+ \, \nu_\ell$ & $\left(0.5 \pm 0.5 \right) \times 10^{-3}$ & $\left(0.5 \pm 0.5 \right) \times 10^{-3}$ \\
        $B \to D^* \pi \pi \, \ell^+ \, \nu_\ell$ & $\left(2.6 \pm 1.0 \right) \times 10^{-3}$ & $\left(2.4 \pm 1.0 \right) \times 10^{-3}$ \\
         $B \to D \eta \, \ell^+ \, \nu_\ell$ & $\left(2.0 \pm 2.0 \right) \times 10^{-3}$ & $\left(2.2 \pm 2.2 \right) \times 10^{-3}$ \\
        $B \to D^{*} \eta \, \ell^+ \, \nu_\ell$ & $\left(2.0 \pm 2.0 \right) \times 10^{-3}$ & $\left(2.2 \pm 2.2 \right) \times 10^{-3}$ \\
         \colrule
         \BtoXclv & $\left(10.8 \pm 0.4\right) \times 10^{-2} $ & $\left(10.0 \pm 0.4\right) \times 10^{-2} $ \\
       \botrule
      \end{tabular}
\end{table}

\section{Event Reconstruction}
\label{sec:reconstruction}
\YFourStoBB events are tagged by fully reconstructing one \PB meson decaying hadronically, also referred to as the tag-side \Btag meson.
The other \PB meson is reconstructed inclusively by selecting a high-momentum lepton. 
The \PX-system is defined by the rest of the event (ROE), consisting of additional unassigned charged particles and neutral clusters in the ECL.
Event-level pre-cuts are applied to reduce the number of continuum and low-multiplicity background components.
We select events with at least four reconstructed charged tracks.
Additionally, we require at least two tracks with $\abs{d_0} < \SI{0.5}{cm}$, $\abs{z_0} < \SI{2}{cm}$ and $\pt > \SI{0.1}{\giga\eVperc}$, as well as at least two ECL clusters with $E > \SI{0.1}{GeV}$ and a polar angle $\theta$ inside the CDC acceptance.
Here, $z_0$ denotes the signed distance in the $z$ direction (parallel to the beams and the magnetic field) of closest approach to the interaction point (POCA). Further, $d_0$ is the signed distance transverse to the $z$ direction to the POCA. To reject continuum events, the event is required to pass $\Rtwo < 0.4$, where $\Rtwo$ is the ratio of the second to the zeroth Fox-Wolfram moment~\cite{Fox:1978vu}.
These event shape variables are calculated using all charged tracks and ECL clusters passing the selection criteria mentioned above.
Finally, the event is required to have a greater visible energy in the CM frame than \SI{4}{GeV}, while the total energy in the ECL is required to lie between $2<E_\mathrm{ECL}<\SI{7}{GeV}$.

\subsection{Hadronic Tag-Side Reconstruction}
The tag-side \Btag candidate is reconstructed using the hadronic tagging algorithm provided by the Full Event Interpretation (FEI) \cite{Keck_2019}.
The FEI uses a fully automated approach to hierarchically reconstruct a tag-side \PB meson and infers a signal probability \FEIprob for each reconstructed \Btag candidate based on multivariate analysis (MVA) techniques.
The algorithm uses an exclusive reconstruction approach resulting in $\mathcal{O}(10'000)$ distinct \PB decay chains.
We use a skimmed version of the data with reconstructed \Btag candidates passing $\FEIprob > 0.001$, $\mbc > \SI{5.24}{\giga\eVperc\squared}$ and $\abs{\DeltaE} < \SI{0.2}{GeV}$.
The beam-constrained mass \mbc and energy difference \DeltaE are defined as
\begin{align}
\mbc &= \sqrt{\frac{s}{4} - (\mathbf{p}^\ast_{\Btag})^2},\\
\DeltaE &= E^\ast_{\Btag} - \frac{\sqrt{s}}{2},
\end{align}
where  $\mathbf{p}^\ast_{\Btag}$ and $E^\ast_{\Btag}$ denote the reconstructed \Btag three-momentum and energy, respectively, in the CM frame.
To further reduce the combinatorial complexity, only the three candidates with the highest FEI signal probability per event for the \Btag candidates are considered in the subsequent stages of the analysis.

\subsection{Selection of Inclusive \BtoXlv Decays}

We select \Pepm, \Pmupm and \PKpm candidates by using the normalized charged particle identification (PID) from sub-detector information.
The \Pepm, \Pmupm and \PKpm candidates are required to have a PID value greater than 0.9, 0.9 and 0.6, respectively.
Additionally, the respective tracks are required to pass $\dr < \SI{1}{cm}$, $|dz| < \SI{2}{cm}$,  have at least one hit in the CDC and a $\theta$ value inside the CDC acceptance.
Here, \dr and $dz$ denote the track's $d_0$ and $z_0$ values, respectively, of its POCA relative to the interaction point. 
To construct the ROE object, we reconstruct all remaining tracks and ECL clusters assuming that they are \Ppipm and photons, respectively.

Electron candidates are corrected for bremsstrahlung by identifying suitable photon candidates. At this stage, the selected light-lepton candidates (\elleqelmu) are combined with the \Btag candidates to form an \PUpsilonFourS candidate.
Due to the fully reconstructed tag-side candidate and the known initial state of the $\APelectron\Pelectron$ collision, the lepton momentum in the signal \PB rest frame, denoted as \plepsigbrestframe, is accessible.
We require lepton candidates with $\plepsigbrestframe > \SI{0.6}{\giga\eVperc}$.
The charge correlations between the \Pqb quark of the \Btag and the signal lepton candidates are not considered when recombining the \PUpsilonFourS candidate, resulting in the eight reconstruction channels $B^+_{\mathrm{tag}} \Pleptonpm$ and $B^0_{\mathrm{tag}} \Pleptonpm$.
In the final analysis, only the $B^+_{\mathrm{tag}} \Pleptonminus$ and $B^0_{\mathrm{tag}} \Pleptonpm$ are considered as signal channels.
The two $B^+_{\mathrm{tag}} \Pleptonplus$ channels are background enriched and used to verify the description of the background modeling.

The hadronic \PX-system is identified from the ROE of the \PUpsilonFourS candidate.
The ROE is constructed using the remaining charged particle and photon candidates that were not used in the reconstruction of the \PUpsilonFourS candidate.
The mass hypothesis of the individual track object is based on the PID selection.
Remaining tracks associated with a kaon likelihood greater than $0.6$ are assigned the kaon mass, while all other ROE tracks are identified as pions.
To remove background candidates that do not belong to the \PUpsilonFourS decay, we consider only tracks in the ROE with $\dr < \SI{2}{cm}$, $\dz < \SI{4}{cm}$, at least one hit in the CDC and a $\theta$ value within the CDC acceptance.
Low-momentum tracks curling inside the CDC are removed prior to construction of the ROE.
Photon candidates are required to pass a region-dependent cut.
We select only photons with $\pt > \SI{20}{MeV}$ and $\mathcal{P}_\mathrm{Zernike}>0.35$,  $\pt > \SI{30}{MeV}$ and $\mathcal{P}_\mathrm{Zernike}>0.15$ and  $\pt > \SI{20}{MeV}$ and $\mathcal{P}_\mathrm{Zernike}>0.4$ for the forward end-cap, barrel, and backward end-cap  ECL region, respectively.
$\mathcal{P}_\mathrm{Zernike}$ denotes the MVA classifier output using Zernike moments~\cite{Hershenhorn:454} of the different clusters.
A second ROE object is constructed with the same selection criteria for the \Btag candidate.
It is used to calculate a set of continuum suppression variables consisting of CLEO cones \cite{PhysRevD.53.1039}, modified Fox-Wolfram moments \cite{Bevan_2014} and thrust information.
These variables are used as input for a boosted decision tree (BDT) to separate \BBpair from continuum events.
We use the BDT algorithm implemented in the \texttt{FastBDT} library \cite{Keck:2017gsv}.

To further reject backgrounds from leptons of secondary decays, misidentified hadrons or continuum events, a cut-based approach is chosen.

Secondary leptons and hadronic fakes are reduced by selecting signal lepton candidates passing $\plepsigbrestframe > \SI{0.8}{\giga\eVperc}$.
To improve the purity of the tag-side reconstruction, we require \Btag candidates with $\FEIprob > 0.01$ and $\mbc > \SI{5.27}{\giga\eVperc\squared}$.
Continuum events are rejected by cutting on the classifier output of the continuum suppression BDT \CSprob.
We select candidates with $\CSprob > 0.7$.
To improve the quality of the reconstructed \PX-system, we require the absolute value of the total charge of the reconstructed event $\totalCharge =Q_{\Btag} + Q_{\Plepton} + Q_{\PX}$ to be less than or equal to one, explicitly allowing a charge imbalance.
Further, the \PX-system is required to contain at least one charged particle.
The missing momentum \pmiss and missing energy \Emiss are required to be greater than $\SI{0.5}{\giga\eVperc}$ and \SI{0.5}{GeV}, respectively.
The absolute value of $\Emiss - \cpmiss$ should be smaller than \SI{0.5}{GeV}.
The missing four-momentum is defined as
\begin{align}
    p_\mathrm{miss}^\mu  =  p_{e^+\,e^-}^\mu -  p_{\Btag}^\mu - p_{\Plepton}^\mu -p_{\PX}^\mu.
\end{align}

\begin{table}[tb]
    \caption{Event selection criteria applied to the reconstructed \PUpsilonFourS candidates.}
\begin{tabular}{lr}
        \toprule
        Variable &  Applied Cut Value  \\
        \colrule
        $\plepsigbrestframe$ &    $>\SI{0.8}{\giga\eVperc}$\\
        $\mbc$  &  $>\SI{5.27}{\giga\eVperc\squared}$ \\
        $\FEIprob$ &   $>0.01$\\
        $\CSprob$  & $> 0.7$\\
        $\abstotalCharge  $  & $\leq1$  \\
        $N_\mathrm{tracks,X}  $ & $\geq 1$  \\
        $\Emiss$  & $>\SI{0.5}{GeV}$  \\
        $\pmiss$   &  $>\SI{0.5}{\giga\eVperc}$\\
        $\absEmissminuscpmiss$   & $<\SI{0.5}{GeV}$ \\
        \botrule
          \label{tab:cutflow}
          \end{tabular}
\end{table}

The event selection criteria are summarized in \cref{tab:cutflow}.
If multiple $\Btag\Plepton$ combinations are present in an event after applying all selection criteria, a best candidate selection (BCS) based on the highest \plepsigbrestframe is performed.
In the case where the same lepton is combined with two different tag-side candidates, the \Btag candidate with the smallest \DeltaE is chosen.

\begin{figure}[tb]
    \centering
    \includegraphics[width=0.75\textwidth]{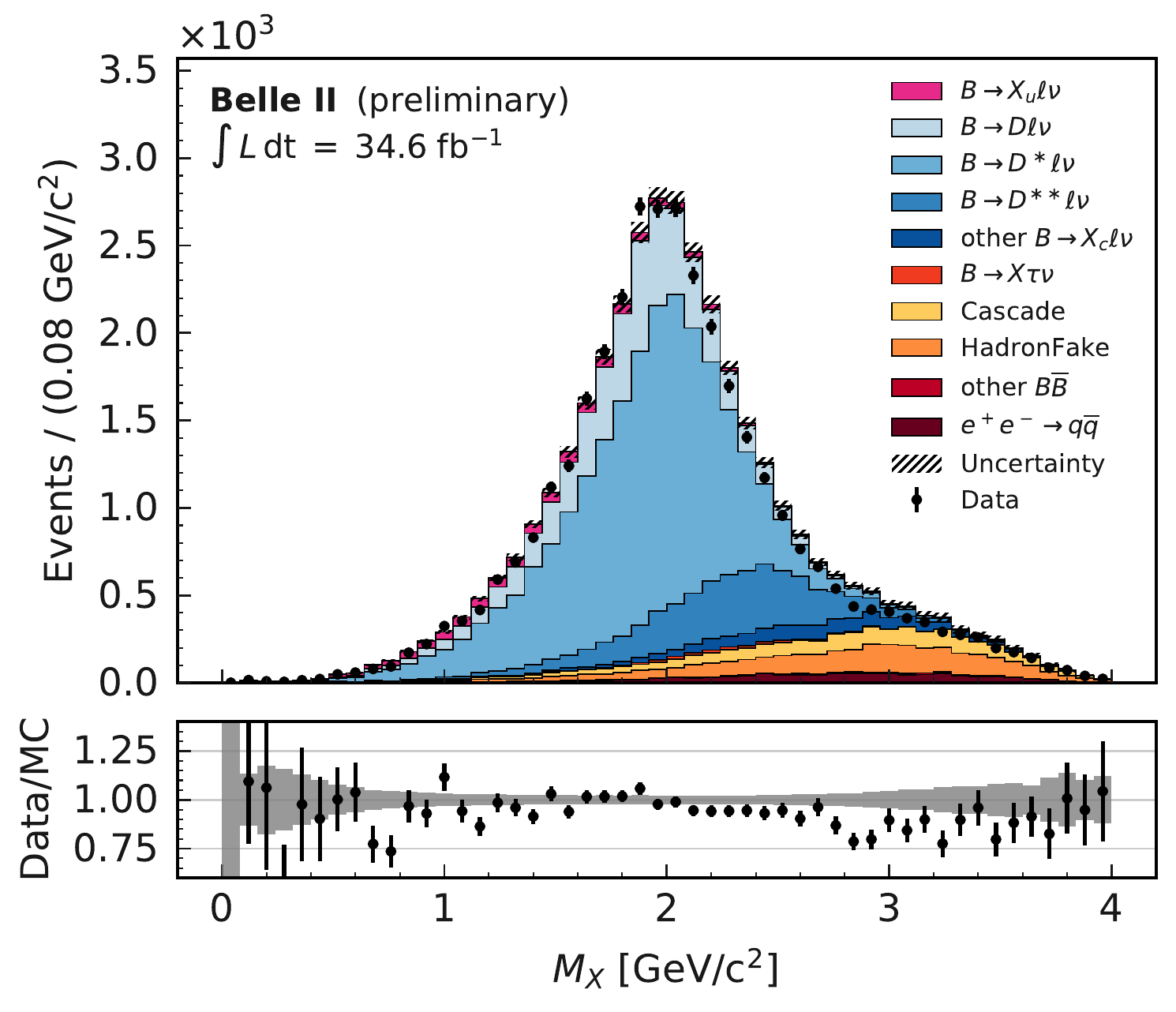}
    \caption{Reconstructed \mx distribution with event selection criteria and BCS applied. The uncertainty band covers the MC statistics, signal lepton PID efficiency and pion fake rate correction, and the FEI efficiency correction for \BBpair and continuum events. At the bottom the per bin ratio of data and MC is shown. The grey boxes display the ratio between the MC expectation taking into account its uncertainty and the nominal value.}
    \label{fig:mx_w_corrections}
\end{figure}

\cref{fig:mx_w_corrections} shows the reconstructed \mx distribution for the full recorded data set with a total integrated luminosity of $\SI{34.6}{fb^{-1}}$.
The displayed MC sample corresponds to an integrated luminosity of $\SI{100}{fb^{-1}}$ and has been scaled to match the luminosity of the recorded data set.
The MC components are corrected for differences in PID and FEI efficiencies between data and simulation.
We correct fake lepton candidates matched to a \Ppi particle on MC level.
The FEI correction factors for the \BBpair components are determined in Ref. \cite{Sutcliffe:1470}, while the correction factors for the continuum component are determined in the side-band of the continuum suppression BDT.

\section{Background Subtraction}
\label{sec:bkg_subtraction}
\begin{figure}
\centering
\includegraphics[width=0.49\textwidth]{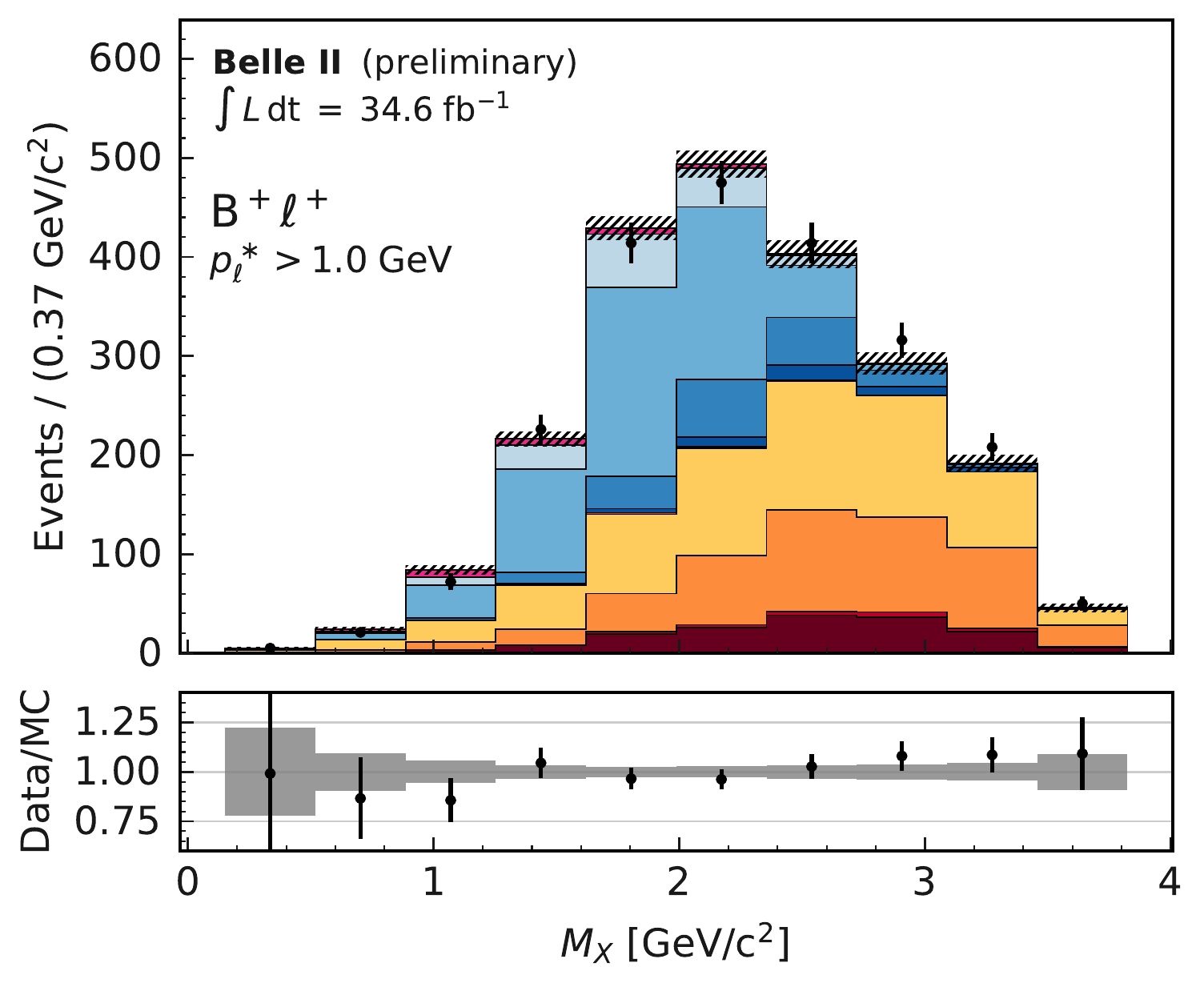}
\includegraphics[width=0.49\textwidth]{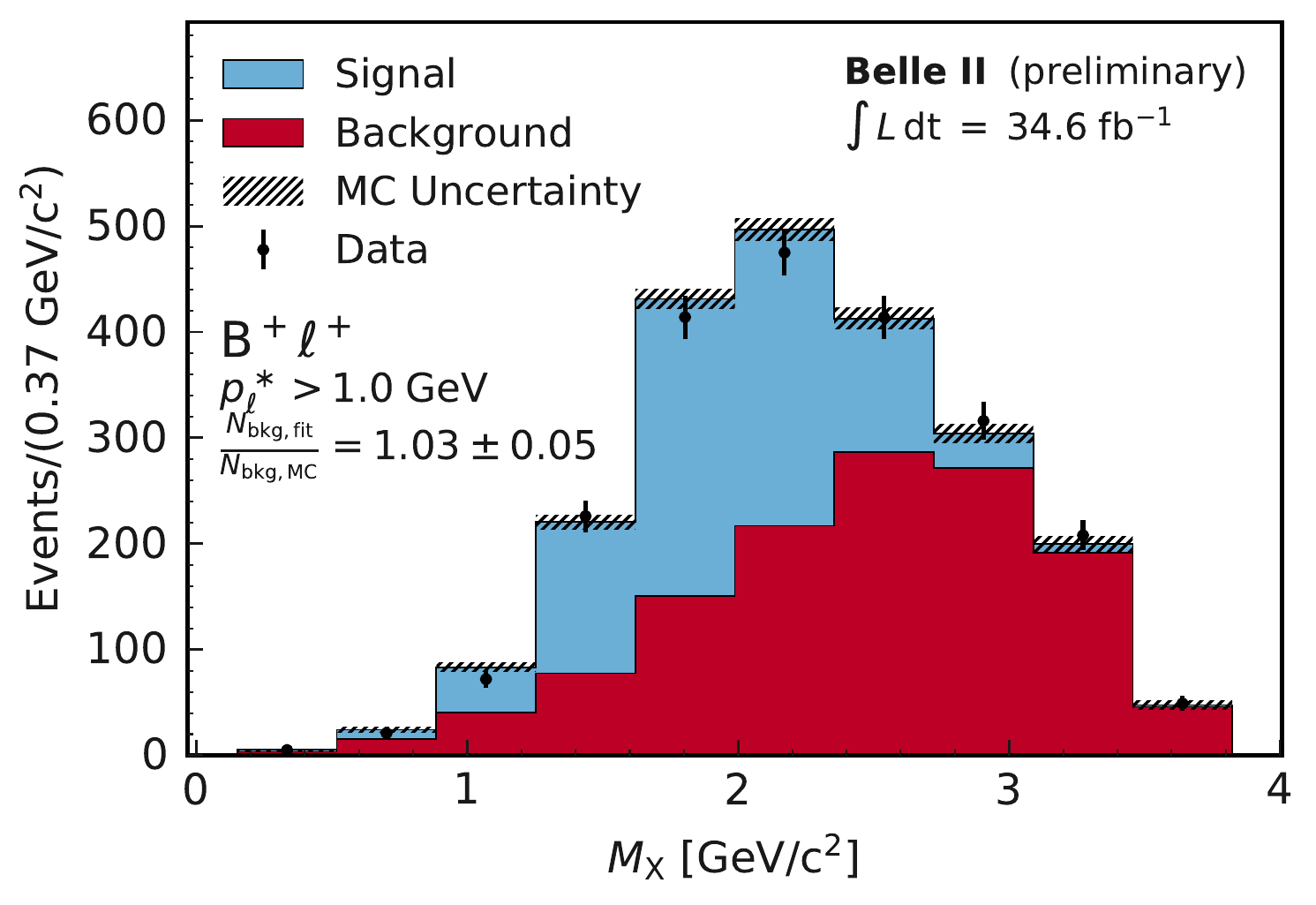} 
\caption{\mx distribution in the $\PBplus \Pleptonplus$ channels for a lower limit of $\plepsigbrestframe > \SI{1.0}{\giga\eVperc}$. The pre-fit \mx spectrum split into sub-components and the post-fit distribution of the two component template fit are shown in the left and right plot, respectively.}
\label{fig:bkg_channel_mx_fit_1}
\end{figure}

The calculation of the hadronic mass moments of \BtoXclv decays requires the subtraction of the remaining background components from the measured events.
To verify the description of the background components in MC, the background enriched reconstruction channels $B^+_{\mathrm{tag}}\Pleptonplus$ are used.
A two component template fit of the \mx distribution is used to determine the number of background events in data.
The background component yield is fitted, while the normalization of the signal template is fixed.
This check is performed for different lower limits on \plepsigbrestframe.
The ratio of the fitted number of background events to the MC expectation is compatible to unity for all lower \plepsigbrestframe cuts.
\cref{fig:bkg_channel_mx_fit_1} shows the pre-fit \mx spectrum split into sub-components in the $B^+_{\mathrm{tag}}\Pleptonplus$ channel for a lower limit on the lepton momentum of $\plepsigbrestframe>\SI{1.0}{\giga\eVperc}$ as well as the post-fit distribution of the signal and background fit.

\begin{figure}
\centering

\includegraphics[width=0.49\textwidth]{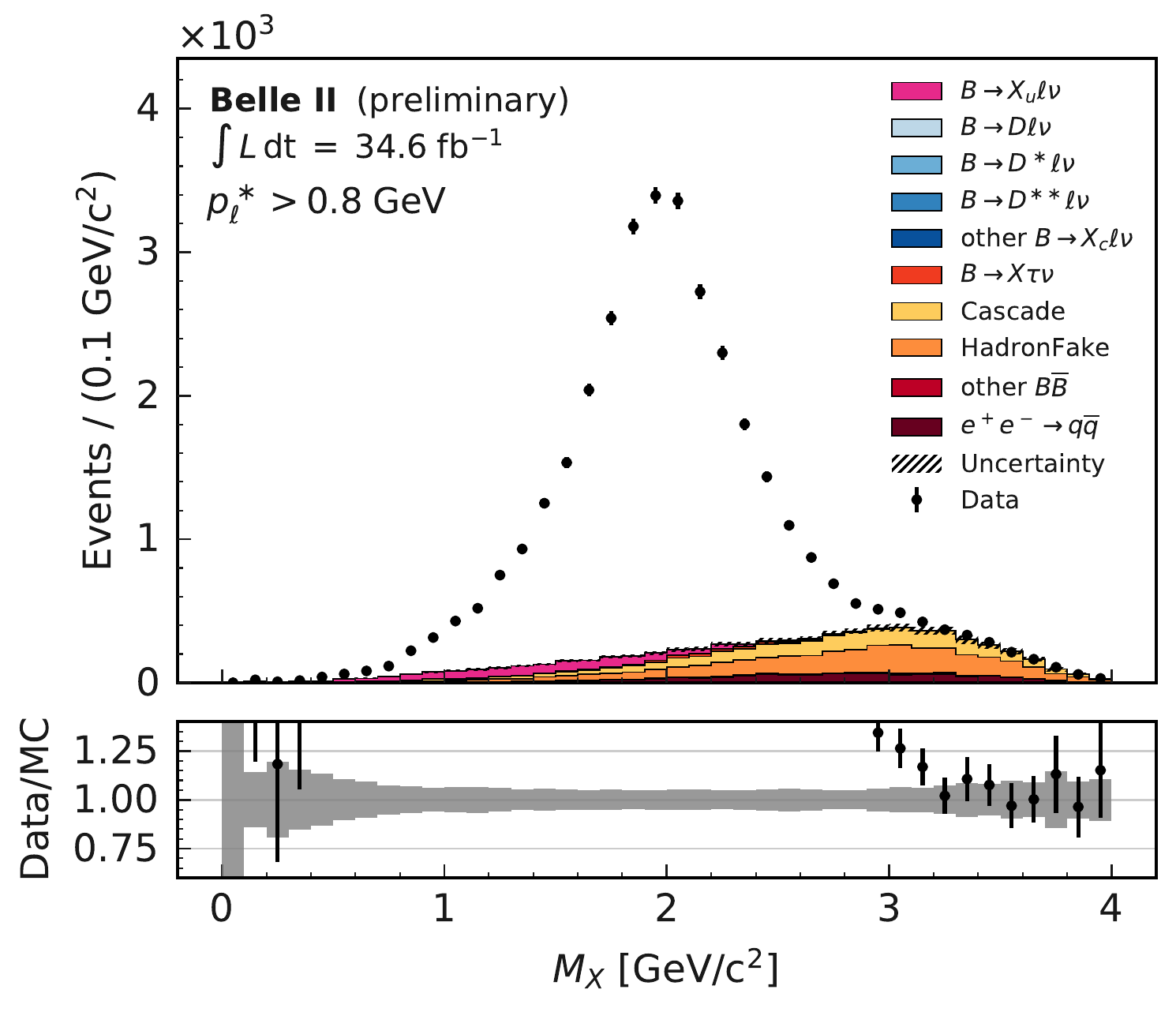} 
\includegraphics[width=0.49\textwidth]{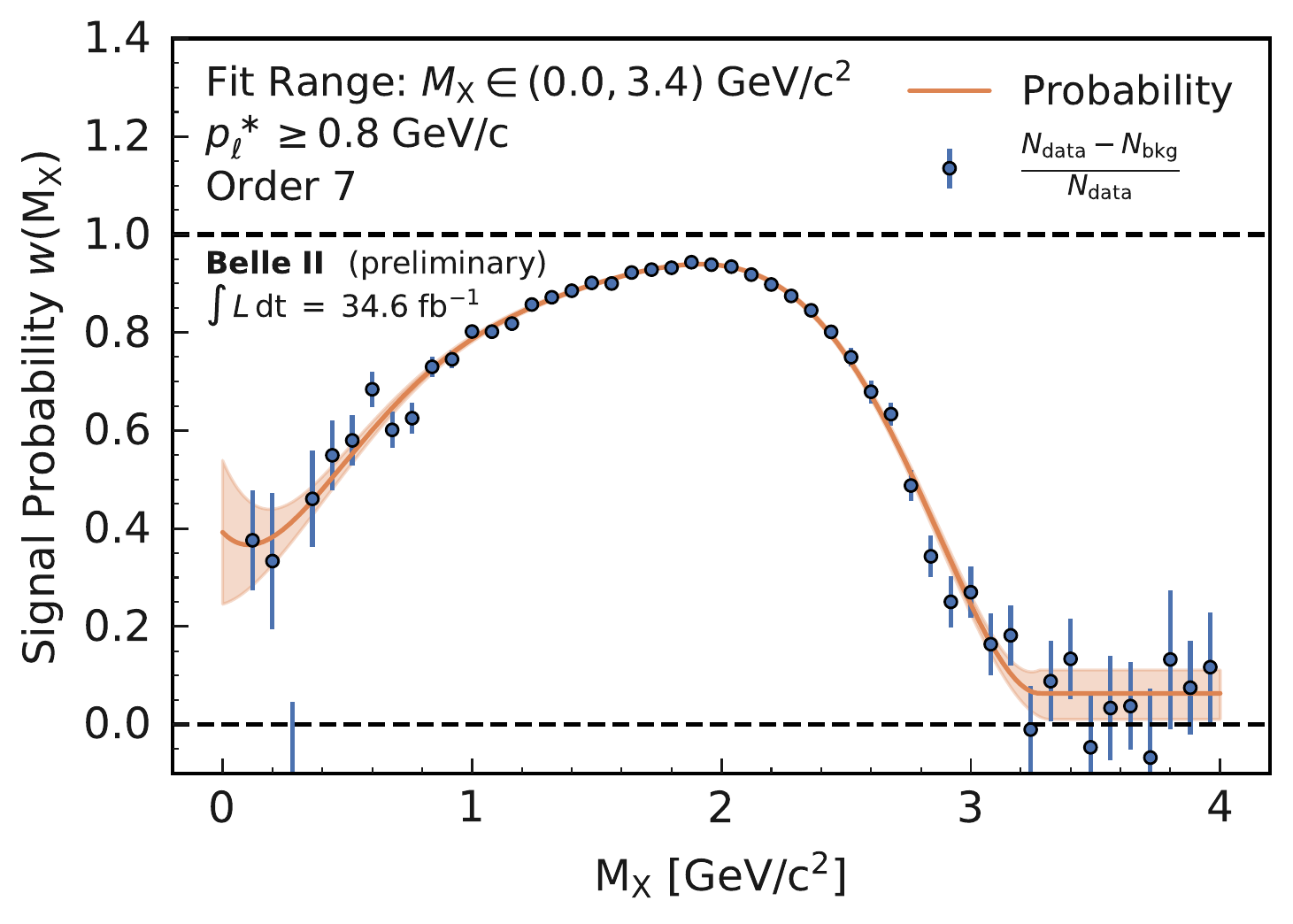} 

\caption{The left column shows the \mx distribution in data and background MC (normalized to the events in data) for $\plepsigbrestframe > \SI{0.8}{\giga\eVperc}$. The corresponding background subtraction factors $w_i$ are shown in the right column together with a fitted Legendre polynomial of degree 7. If the fit has a minimum at the left or right tail, the polynomial is replaced with a constant value. The uncertainties are from statistical uncertainties only.}
\label{fig:bkg_poly_fit_1}
\end{figure}

We subtract the background by assigning a signal probability to each event.
The signal probability $w_i(\mx)$ is determined from a fit of the bin-wise difference between the measured \mx spectrum and the remaining background MC components normalized to the measured distribution
\begin{align}
\label{eq:background_subtraction_normalized_bin_wise}
    w_i(\mx)=\frac{N^\mathrm{data}_i - N^\mathrm{bkg,MC}_i}{N^\mathrm{data}_i},
\end{align}
where the index $i$ denotes the corresponding \mx bin.
To get a continuous description of the signal probability, we fit a series of Legendre polynomials to the bin-wise probabilities.
Prior to fitting, the fit-range is transformed to the interval $[-1, 1]$ to exploit the orthogonal nature of the polynomials.
The order of the Legendre polynomial is determined by cutting off the series when the next higher order fitted coefficient is compatible with zero.
If the fit reaches a minimum in the background dominated low or high hadronic mass values, the polynomial is replaced by a constant value equal to the found minimum. 
The procedure is performed for different lower limits on the lepton momentum \plepsigbrestframe.
\cref{fig:bkg_poly_fit_1} shows the fitted signal probability as a function of the reconstructed \mx with $\plepsigbrestframe>\SI{0.8}{\giga\eVperc}$ and the measured \mx spectrum compared to the background MC components.

\section{Measurement of Hadronic Mass Moments}
\label{sec:mx_moments_measurements}
\subsection{Extraction of Moments}
\label{subesc:calibration}
\begin{figure}[tb]
\centering
\includegraphics[width=0.49\textwidth]{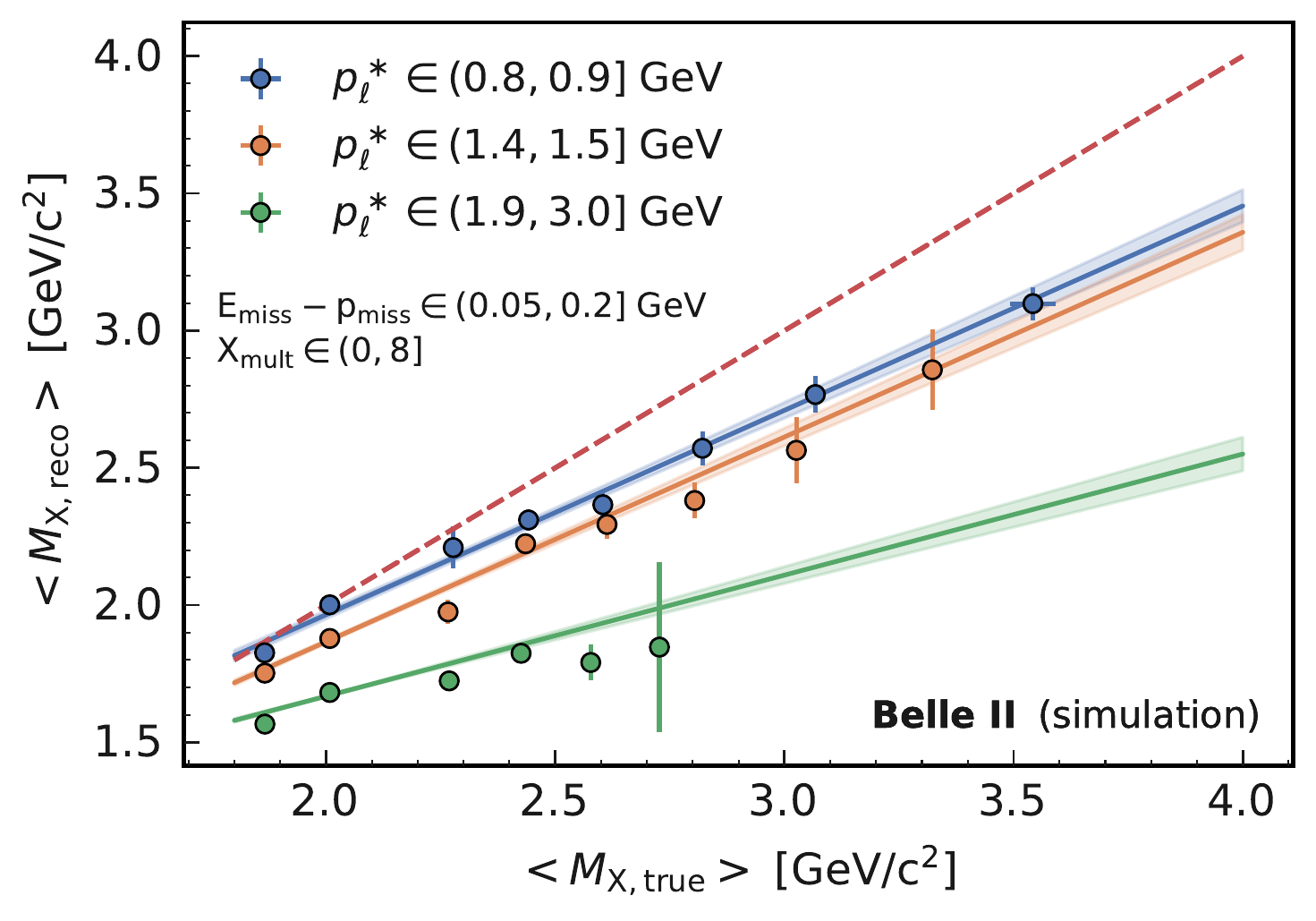}
\caption{Example of the calibration curves for the first moment \mxmoment{} in bins of \Emissminuspmiss, \Xmultiplicity and \plepsigbrestframe. The moments \mxmomentreco{} versus \mxmomenttrue{} calculated in bins of \mxtrue are shown. The uncertainty of the calibration curves takes into account the statistical uncertainty on the fitted slope and intercept. The red dashed reference line shows $\mxmomenttrue{} = \mxmomentreco{}$}
\label{fig:mx_calibration_curves}
\end{figure}

\begin{figure}[tb]
\centering
\includegraphics[width=0.49\textwidth]{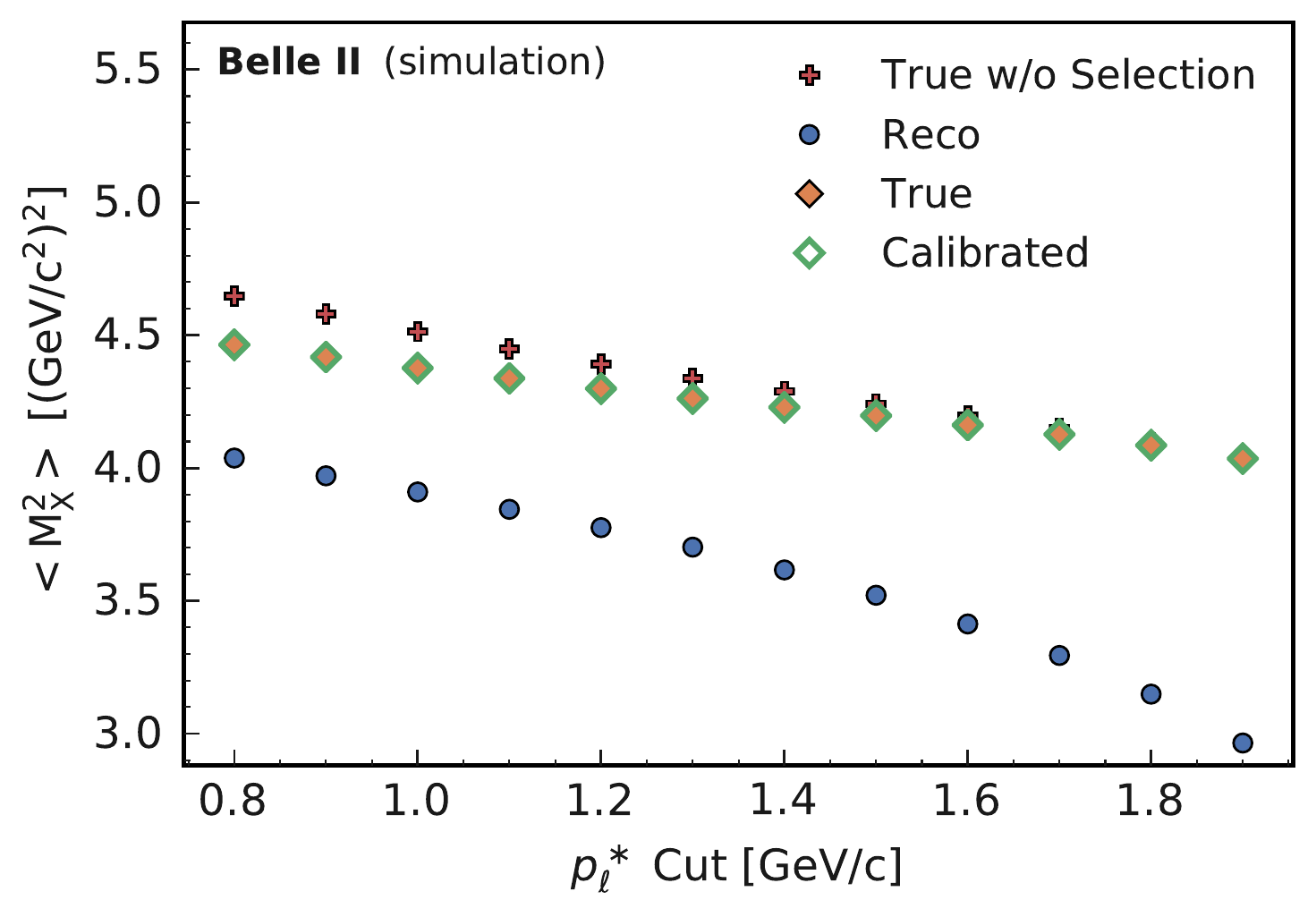}
\caption{Second hadronic mass moment \mxmoment{2} calculated on signal MC for different lower limits on \plepsigbrestframe. The plotted moments are the measured uncalibrated, calibrated and true moments after the application of all analysis selection criteria. In addition, the true \mxmoment{2} calculated from the MC sample without any selection criteria applied are shown as red crosses.}
\label{fig:mx_calibrated_cut}
\end{figure}

To extract unbiased moments, the measured $\mx^n$ spectrum has to be corrected for effects that distort the measured distribution.
We derive calibration functions based on MC simulation to describe the relationship between the reconstructed moments \mxmomentreco{n} and the moments calculated at the generator level \mxmomenttrue{n}.
Both moments are calculated in bins of the generator level $M_\mathrm{X}^n$ distribution.
We find a linear relationship between \mxmomentreco{n} and \mxmomenttrue{n}, which allows us to calculate a calibrated \mx value
\begin{align}
    \mnxcalib{n} = \frac{\mnx{n} - c(\Emissminuspmiss, \Xmultiplicity, \plepsigbrestframe)}{m(\Emissminuspmiss, \Xmultiplicity, \plepsigbrestframe)} \, .
\end{align}
Here $c$ and $m$ denote the fitted intercept and slope of the linear calibration functions, respectively.
Since the bias of the measured \mx spectrum is not constant over the available phase-space, the calibration is performed in bins of \plepsigbrestframe, \Emissminuspmiss, and the particle multiplicity of the \PX-system denoted as \Xmultiplicity.
We use bins in \plepsigbrestframe with a width of $\SI{0.1}{\giga\eVperc}$ between $0.8$ and \SI{1.9}{\giga\eVperc} and one bin for $\plepsigbrestframe\geq \SI{1.9}{\giga\eVperc}$.
A binning of $[-0.5, 0.05, 0.2, 0.5]\;\mathrm{GeV}$ and $[1,8, 30]$ is used for \Emissminuscpmiss and \Xmultiplicity, respectively.
Due to limited statistics in the phase space above $\plepsigbrestframe \geq \SI{1.7}{\giga\eVperc}$, the additional binning in \Emissminuscpmiss and \Xmultiplicity is not used in this region.
\cref{fig:mx_calibration_curves} shows an example of three calibration curves for \mxmoment{} in three bins of \plepsigbrestframe and one bin in \Emissminuscpmiss and \Xmultiplicity.
\cref{fig:mx_calibrated_cut} shows the second hadronic mass moment \mxmoment{2} from signal MC before and after the application of the calibration procedure.
The second moments of the \BtoXclv MC at generator level with and without the application of event selection criteria are also shown.

Together with the signal probability $w_i$ and the calibrated \mxcalib distribution, the \mxmoments can be calculated without unfolding the measured \mx spectrum.
The hadronic mass moments are calculated as a weighted average using
\begin{align}
    \mxmoments = \frac{\sum_{i} w_i(\mx) \mxcalib {}^n_i}{\sum_{i} w_i(\mx)} \times \mathcal{C}_\mathrm{calib} \times \mathcal{C}_\mathrm{true}.
    \label{eq:moment_calculation}
\end{align}
The two additional factors \ccalib and \ctrue correct a remaining bias due to the calibration and selection efficiencies for different \BtoXclv components.
The factor $\ccalib=~\mxmomentstrue /\mxmomentscalib $ corrects the remaining bias of the calibrated moments and the true moments for each lower limit on \plepsigbrestframe.
We observe remaining bias corrections ranging between $1.001$ for the first moment up to $0.988$ for the fourth moment.
To correct a possible bias due to the event selection criteria applied, we apply a second correction factor $\ctrue = \mxmomentstruesignal/\mxmomentstrue$.
Here, \mxmomentstruesignal are the moments of the generator \mx spectrum of our simulated \BtoXclv decays without the application of the aforementioned event selection criteria.
Only a cut on the generator level lepton momentum in the signal \PB meson rest frame is applied.
To be able to correct for the effect of final state radiation on the lepton momentum, the MC sample used to calculate \mxmomentstruesignal does not include the simulation of radiative photons with \texttt{PHOTOS}.
We obtain values for \ctrue ranging from $1.02$ to $1.27$ for the lowest \plepsigbrestframe cut. 
For higher \plepsigbrestframe cuts the \ctrue ranges from $1.00$ to $1.01$ for the highest cut value.

\subsection{Uncertainties}
\label{subsec:uncertainties}
We identify several sources of statistical and systematic uncertainties. The total uncertainty is calculated by adding statistical and systematic uncertainties in quadrature.

For the statistical uncertainty, we consider two different components.
The \mxmoments are calculated as a weighted mean over all events.
We calculate the variance of the weighted mean as~\cite{GATZ19951185} 
\begin{align}
    V({\mxmoments}) = \frac{n}{(n-1)\sum_i^n w_i}\sum_i^n w_i^2(\mnxcalibevent{n} - \mxmoments)^2.
\end{align}
We verifie the validity of this formula applying a bootstrapping approach.
The second part of the statistical uncertainty is given by the statistical uncertainty of the polynomial coefficients of the signal probability function.
The uncertainty is propagated by using error propagation to calculate the uncertainty on the signal probability.
To estimate the impact of the propagated uncertainty on the measured \mxmoments, the calculation of the moments is repeated with varied signal probability values.
The total statistical uncertainty is calculated by summing both uncertainties in quadrature.

To estimate the impact of systematic uncertainties, the following effects are taken into account:
\begin{enumerate}
   \item Statistical uncertainty on the linear calibration functions:
    \newline
    The used linear calibration functions are determined using a dedicated MC sample of \BtoXclv decays.
    Both the slope and the intercept have statistical uncertainties and are correlated.
    To propagate the uncertainties correctly with their correlations, the eigenvalues and eigenvectors of the covariance  matrix are used to calculate two orthogonal variations of both parameters via
    \begin{align}
        c_i^\pm = c_i^{nom} \pm \sqrt{\lambda_i} \hat{e}_i,
    \end{align}
    where $c_i^{nom}$ and $c_i^\pm$ denote the nominal and varied parameters, respectively, of the linear calibration function.
    $\lambda_i$ and $\hat{e}_i$ are the $i$-th eigenvalue and eigenvector of the parameter covariance matrix.
    In total, we get two ($i=1,2$) independent variations of the determined parameters.
    
    The impact of these uncertainties is estimated by repeating the calculation of the \mx moments and taking the total value of the difference of each variation divided by two as a source of uncertainty. A larger set of MC events would reduce this systematic.
    
    \item FEI and PID efficiency correction uncertainty:
    \newline
    The FEI efficiency correction uncertainty is propagated by varying the efficiency correction by its uncertainty and repeating the determination of the background subtraction weights.
    Again, the uncertainty is taken as half of the total value of the resulting difference of \mxmoments calculated with varied probabilities.
    
    The PID uncertainty is estimated using the set of varied nominal weights in bins of \mx. 
    The PID correction for each event is varied by the estimated bin-wise uncertainty.
    To gauge the impact of this source of uncertainty, the same method as for the FEI efficiency uncertainty determination is used.
    
    \item \BtoXulv branching fraction uncertainty:
    \newline
    The \BtoXulv branching fraction uncertainty is estimated to be 14\% using the latest experimental average of $\SI{2.13\pm0.30}{\%}$ \cite{PhysRevD.98.030001}.
    The corresponding MC component is varied accordingly and the signal probability function is redetermined using the varied MC sample.
    
    \item Statistical uncertainty on the bias correction factor $\ccalib \times \ctrue$:
    \newline
    The remaining bias correction also contains a statistical uncertainty due to the limited number of MC events used to determine it.
    The \mx moments are calculated by varying the bias correction factor according to this statistical uncertainty. 
    
    \item Composition of higher mass \PXc states: 
    \newline
    The bias correction factor \ctrue yields a significant correction to the final result.
    The origin of this correction is the underlying modeling of the higher mass states of the \BtoXclv spectrum, which has changed in comparison to previous analyses.
    The uncertainty of this correction factor is determined by assigning a $100\%$ uncertainty to the branching fraction of the non-resonant part of the \PXc spectrum and repeating the calculation for \ctrue.
    The $100\%$ uncertainty on the non-resonant \BtoXclv branching fractions is a conservative choice, since the decays contributing to this region of the spectrum are not determined experimentally.
    The resulting uncertainty is propagated to the \mxmoments values by repeating the calculation with the varied \ctrue and taking the absolute value of the difference to the nominal moments as the systematic uncertainty.
 
\end{enumerate}
To estimate the total systematic uncertainty, all considered sources of systematics are added in quadrature.

\subsection{Results}
\label{subsec:results}
\begin{figure}
    \centering
    \includegraphics[width=0.49\textwidth]{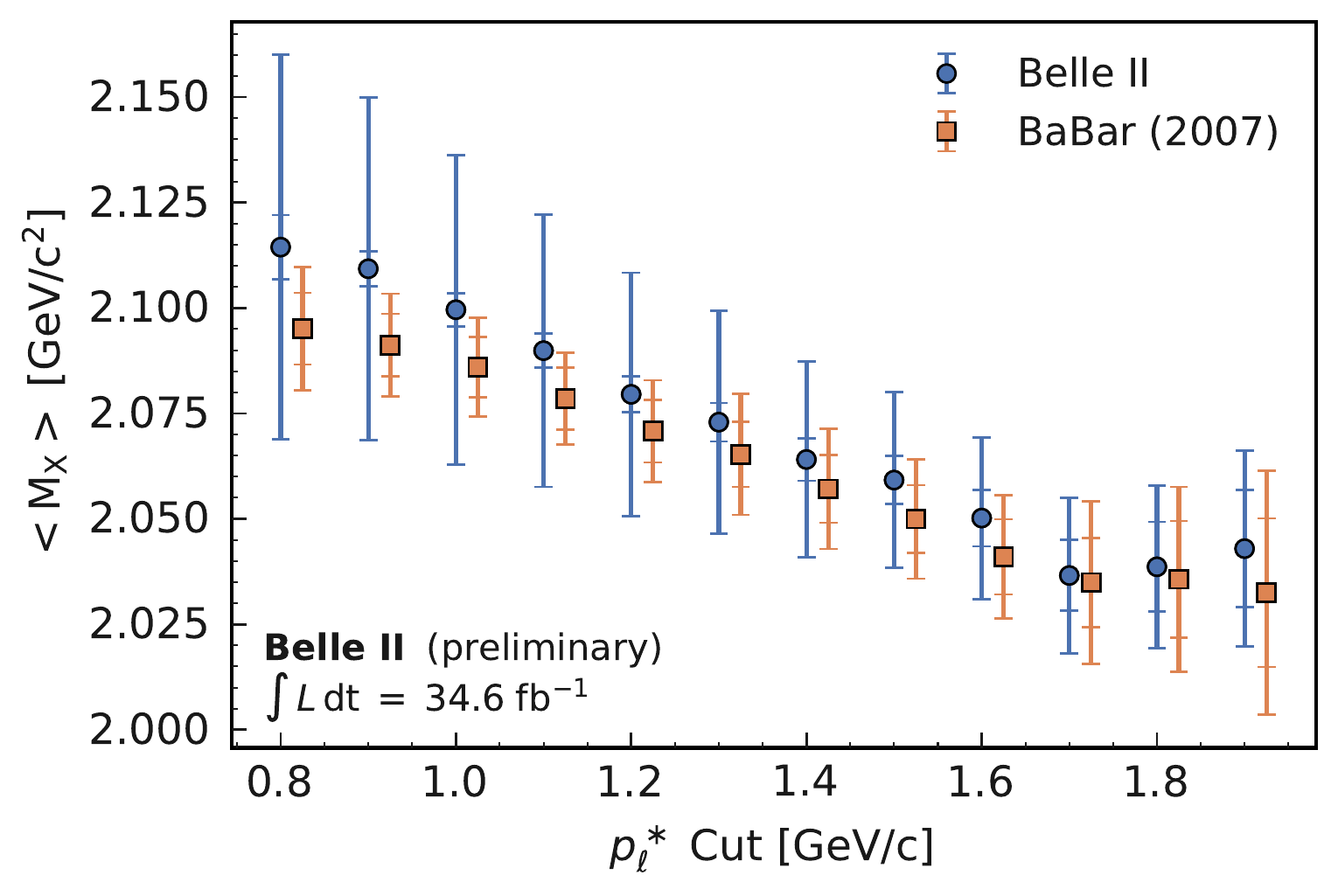}
    \includegraphics[width=0.49\textwidth]{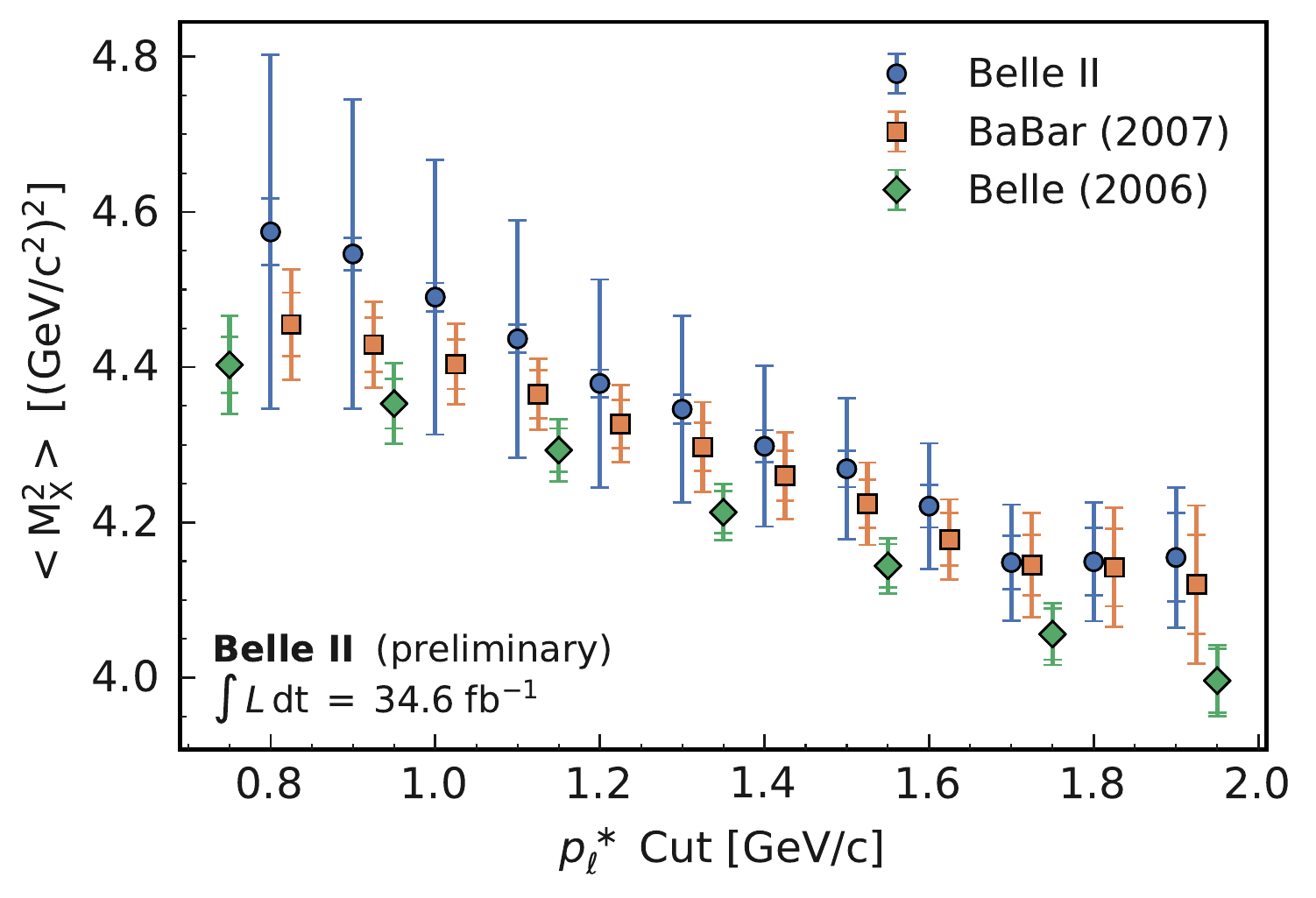}\\
    \includegraphics[width=0.49\textwidth]{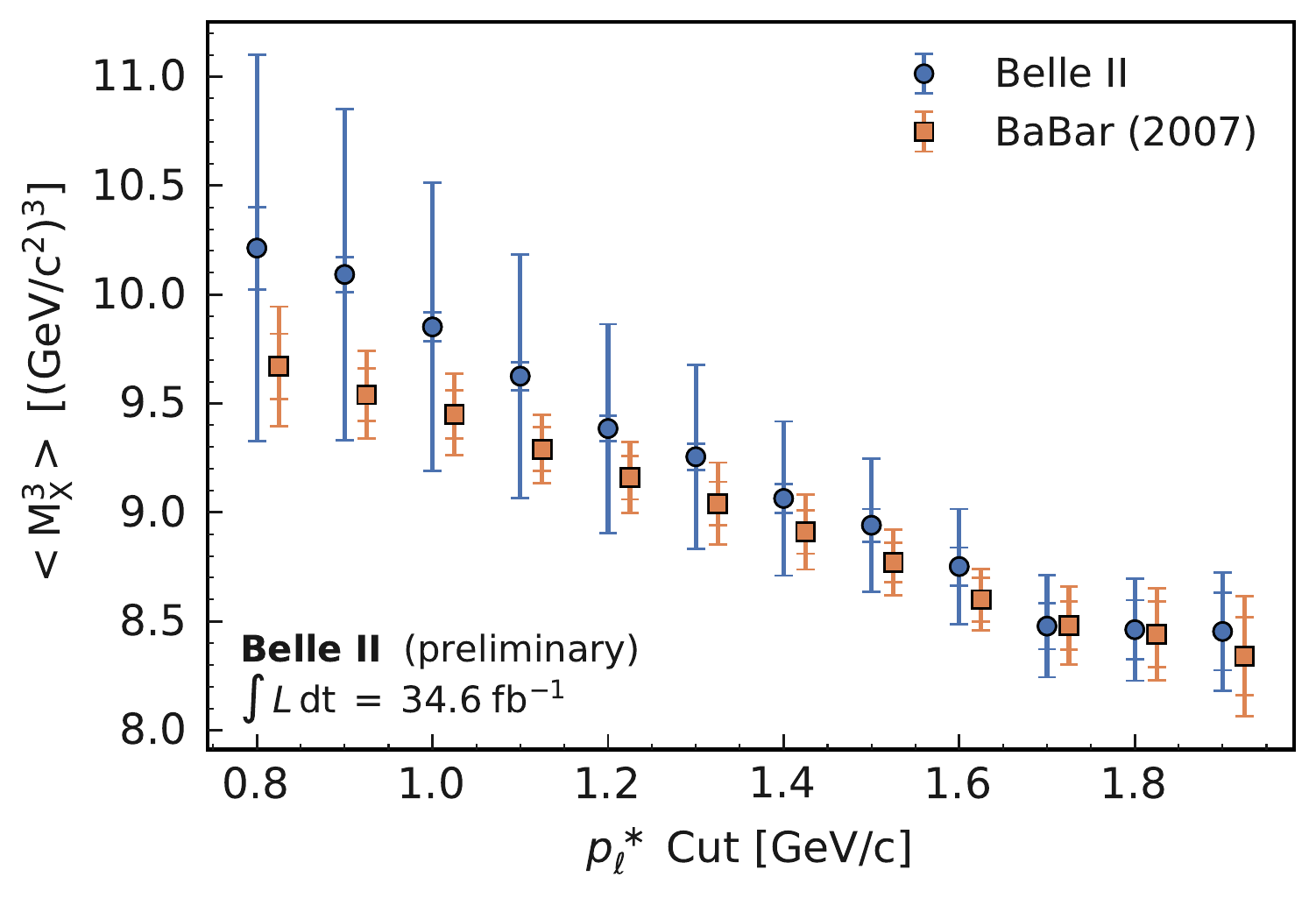}
    \includegraphics[width=0.49\textwidth]{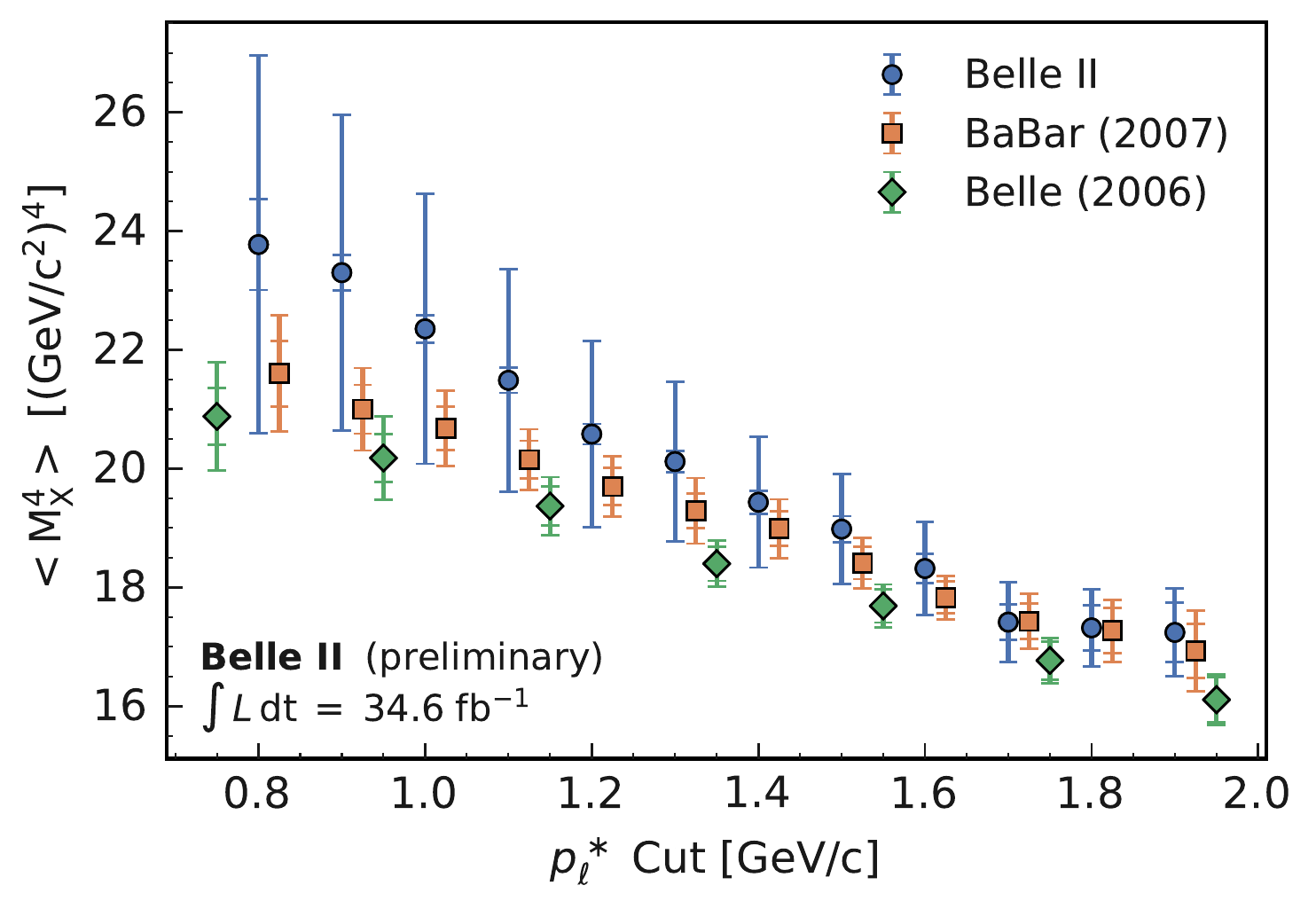}\\
    \includegraphics[width=0.49\textwidth]{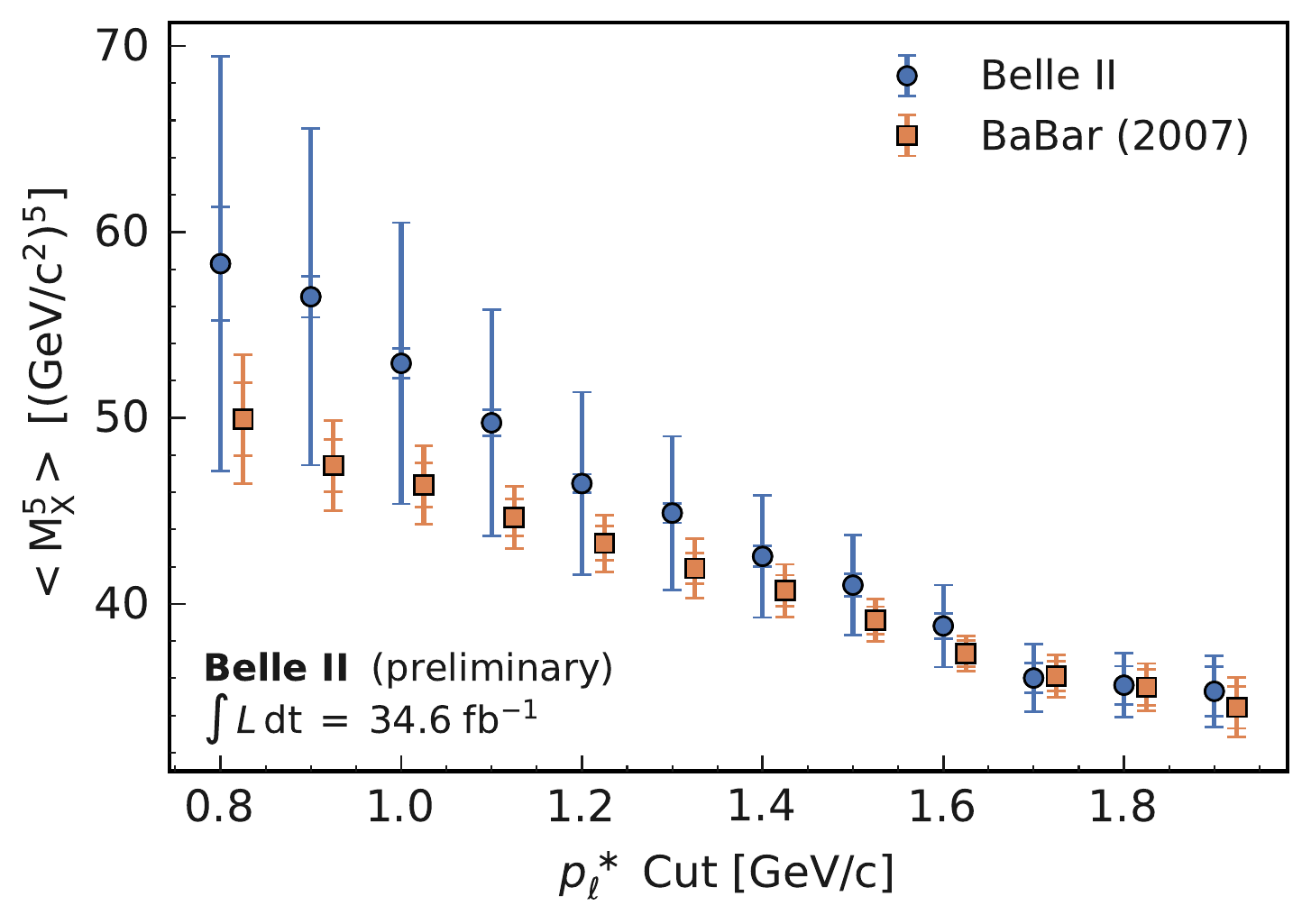}
    \includegraphics[width=0.49\textwidth]{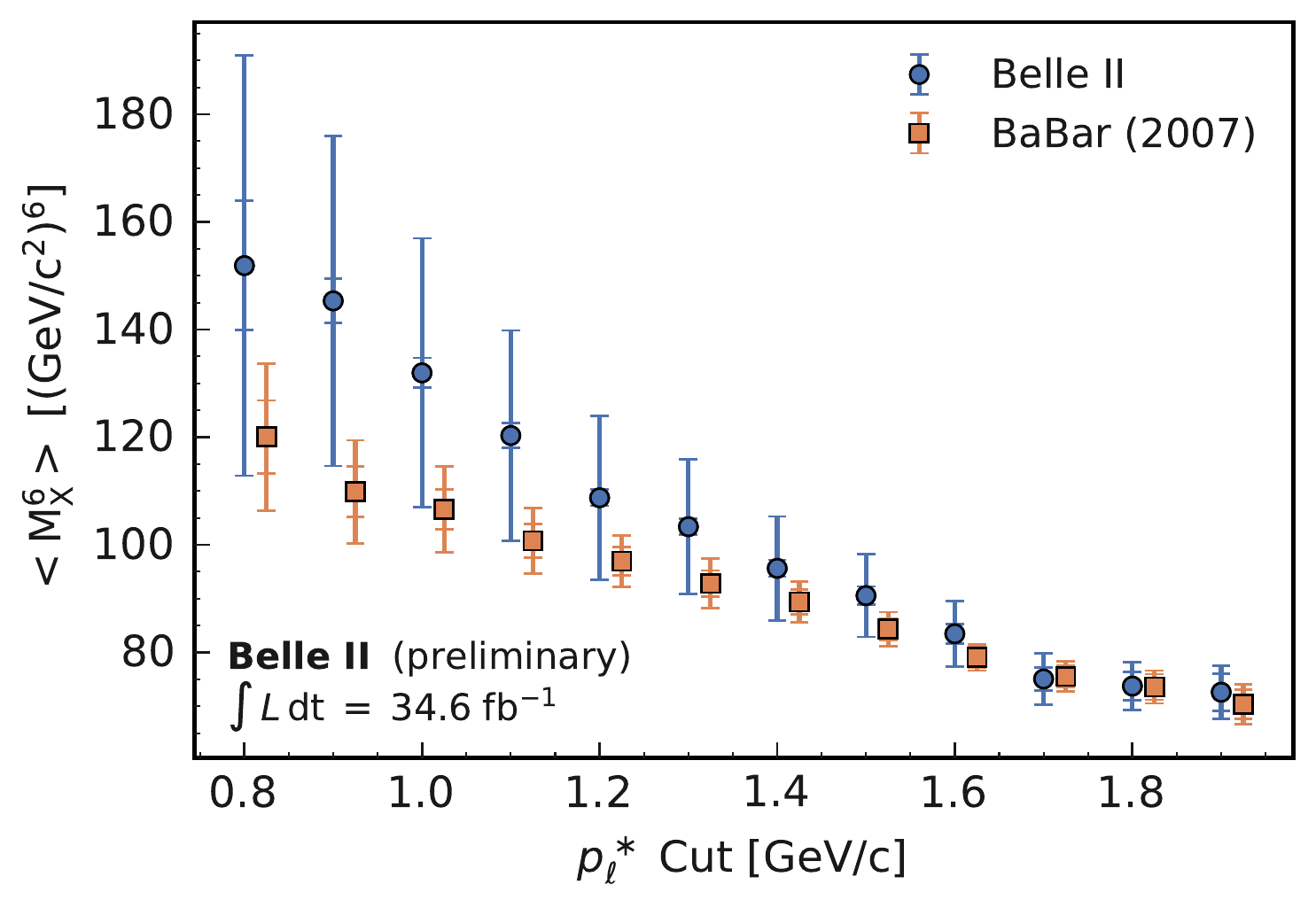}
    
    \caption{Measured \mxmoments moments as a function of different \plepsigbrestframe cuts. The error-bars correspond to the statistical (inner) and total (outer) uncertainty calculated by adding the statistical and systematic error in quadrature. A comparison to previous \mxmoments measurements from BaBar (2007) and Belle (2006) is shown as reference points. The current precision is not yet competitive with the previous results.}
    \label{fig:mx_moments_on_data}
\end{figure}

The measured hadronic mass moments are shown in \cref{fig:mx_moments_on_data} as a function of a lower limit on the lepton momentum in the signal \PB rest frame.
The results of previous analyses performed by BaBar \cite{Aubert:2007yaa} and Belle \cite{Schwanda_2007} are shown for comparison.
The results agree within the uncertainties, but the current precision is not yet competitive.
The numerical values, together with the itemization of the full statistical and systematic uncertainties, are given in \cref{sec:numerical_values}.
The measured moments show a clear dependence on the \plepsigbrestframe cut, resulting in smaller \mxmoments values for higher \plepsigbrestframe cuts.
The uncertainties of the moments for lower \plepsigbrestframe cuts are dominated by the systematic components, while those for higher \plepsigbrestframe cuts have a higher statistical uncertainty.

\section{Summary}
\label{sec:summary}
We have presented a preliminary measurement the first six moments of the hadronic mass spectrum in \BtoXclv decays.
The \mxmoments are measured as a function of a lower cut on the lepton momentum in the signal \PB rest frame \plepsigbrestframe.
The results agree with previous measurements within their uncertainties, but tend to higher nominal values for lower cuts on \plepsigbrestframe.
The moments are calculated as a weighted mean using signal probabilities as event-wise weights.
The achieved precision is not yet competitive with previous analyses.
The systematic uncertainties, in particular, can decrease in futures measurements by reducing the bias in the reconstructed \mx distribution as well as more extensive studies on the composition of unmeasured parts of the \BtoXclv spectrum.
\section*{Acknowledgements}

We thank the SuperKEKB group for the excellent operation of the
accelerator; the KEK cryogenics group for the efficient
operation of the solenoid; and the KEK computer group for
on-site computing support.
This work was supported by the following funding sources:
Science Committee of the Republic of Armenia Grant No. 18T-1C180;
Australian Research Council and research grant Nos.
DP180102629, 
DP170102389, 
DP170102204, 
DP150103061, 
FT130100303, 
and
FT130100018; 
Austrian Federal Ministry of Education, Science and Research, and
Austrian Science Fund No. P 31361-N36; 
Natural Sciences and Engineering Research Council of Canada, Compute Canada and CANARIE;
Chinese Academy of Sciences and research grant No. QYZDJ-SSW-SLH011,
National Natural Science Foundation of China and research grant Nos.
11521505,
11575017,
11675166,
11761141009,
11705209,
and
11975076,
LiaoNing Revitalization Talents Program under contract No. XLYC1807135,
Shanghai Municipal Science and Technology Committee under contract No. 19ZR1403000,
Shanghai Pujiang Program under Grant No. 18PJ1401000,
and the CAS Center for Excellence in Particle Physics (CCEPP);
the Ministry of Education, Youth and Sports of the Czech Republic under Contract No.~LTT17020 and 
Charles University grants SVV 260448 and GAUK 404316;
European Research Council, 7th Framework PIEF-GA-2013-622527, 
Horizon 2020 Marie Sklodowska-Curie grant agreement No. 700525 `NIOBE,' 
and
Horizon 2020 Marie Sklodowska-Curie RISE project JENNIFER2 grant agreement No. 822070 (European grants);
L'Institut National de Physique Nucl\'{e}aire et de Physique des Particules (IN2P3) du CNRS (France);
BMBF, DFG, HGF, MPG, AvH Foundation, and Deutsche Forschungsgemeinschaft (DFG) under Germany's Excellence Strategy -- EXC2121 ``Quantum Universe''' -- 390833306 (Germany);
Department of Atomic Energy and Department of Science and Technology (India);
Israel Science Foundation grant No. 2476/17
and
United States-Israel Binational Science Foundation grant No. 2016113;
Istituto Nazionale di Fisica Nucleare and the research grants BELLE2;
Japan Society for the Promotion of Science,  Grant-in-Aid for Scientific Research grant Nos.
16H03968, 
16H03993, 
16H06492,
16K05323, 
17H01133, 
17H05405, 
18K03621, 
18H03710, 
18H05226,
19H00682, 
26220706,
and
26400255,
the National Institute of Informatics, and Science Information NETwork 5 (SINET5), 
and
the Ministry of Education, Culture, Sports, Science, and Technology (MEXT) of Japan;  
National Research Foundation (NRF) of Korea Grant Nos.
2016R1\-D1A1B\-01010135,
2016R1\-D1A1B\-02012900,
2018R1\-A2B\-3003643,
2018R1\-A6A1A\-06024970,
2018R1\-D1A1B\-07047294,
2019K1\-A3A7A\-09033840,
and
2019R1\-I1A3A\-01058933,
Radiation Science Research Institute,
Foreign Large-size Research Facility Application Supporting project,
the Global Science Experimental Data Hub Center of the Korea Institute of Science and Technology Information
and
KREONET/GLORIAD;
Universiti Malaya RU grant, Akademi Sains Malaysia and Ministry of Education Malaysia;
Frontiers of Science Program contracts
FOINS-296,
CB-221329,
CB-236394,
CB-254409,
and
CB-180023, and SEP-CINVESTAV research grant 237 (Mexico);
the Polish Ministry of Science and Higher Education and the National Science Center;
the Ministry of Science and Higher Education of the Russian Federation,
Agreement 14.W03.31.0026;
University of Tabuk research grants
S-1440-0321, S-0256-1438, and S-0280-1439 (Saudi Arabia);
Slovenian Research Agency and research grant Nos.
J1-9124
and
P1-0135; 
Agencia Estatal de Investigacion, Spain grant Nos.
FPA2014-55613-P
and
FPA2017-84445-P,
and
CIDEGENT/2018/020 of Generalitat Valenciana;
Ministry of Science and Technology and research grant Nos.
MOST106-2112-M-002-005-MY3
and
MOST107-2119-M-002-035-MY3, 
and the Ministry of Education (Taiwan);
Thailand Center of Excellence in Physics;
TUBITAK ULAKBIM (Turkey);
Ministry of Education and Science of Ukraine;
the US National Science Foundation and research grant Nos.
PHY-1807007 
and
PHY-1913789, 
and the US Department of Energy and research grant Nos.
DE-AC06-76RLO1830, 
DE-SC0007983, 
DE-SC0009824, 
DE-SC0009973, 
DE-SC0010073, 
DE-SC0010118, 
DE-SC0010504, 
DE-SC0011784, 
DE-SC0012704; 
and
the National Foundation for Science and Technology Development (NAFOSTED) 
of Vietnam under contract No 103.99-2018.45.

\bibliography{belle2-note-template}

\providecommand{\href}[2]{#2}\begingroup\raggedright\begin{thebibliography}{10}

\bibitem{Gambino:2020jvv}
P.~Gambino {\em et~al.}, ``{Challenges in Semileptonic \PB Decays}'',  (6,
  2020) , \href{http://arxiv.org/abs/2006.07287}{{\ttfamily arXiv:2006.07287}}.

\bibitem{Keck_2019}
T.~Keck {\em et~al.}, ``{The Full Event Interpretation}'',
  \href{http://dx.doi.org/10.1007/s41781-019-0021-8}{{\em Computing and
  Software for Big Science} {\bfseries 3} no.~1, (Feb, 2019) }.
  \url{http://dx.doi.org/10.1007/s41781-019-0021-8}.

\bibitem{Abe:2010sj}
{\bfseries Belle II}, T.~Abe, ``{Belle II Technical Design Report}'',  (2010) ,
\href{http://arxiv.org/abs/1011.0352}{{\ttfamily arXiv:1011.0352}}.

\bibitem{Akai:2018mbz}
K.~Akai {\em et~al.}, ``{SuperKEKB Collider}'',
  \href{http://dx.doi.org/10.1016/j.nima.2018.08.017}{{\em Nucl. Instrum. Meth.
  A} {\bfseries 907} (2018) 188--199},
  \href{http://arxiv.org/abs/1809.01958}{{\ttfamily arXiv:1809.01958}}.

\bibitem{Kou:2018nap}
{\bfseries Belle II}, W.~Altmannshofer {\em et~al.}, ``{The Belle II Physics
  Book}'', \href{http://dx.doi.org/10.1093/ptep/ptz106}{{\em PTEP} {\bfseries
  2019} no.~12, (2019) 123C01},
  \href{http://arxiv.org/abs/1808.10567}{{\ttfamily arXiv:1808.10567}}.
  [Erratum: PTEP 2020, 029201 (2020)].

\bibitem{LANGE2001152}
D.~Lange, ``{The EvtGen particle decay simulation package}'',
  \href{http://dx.doi.org/https://doi.org/10.1016/S0168-9002(01)00089-4}{{\em
  Nuclear Instruments and Methods in Physics Research Section A: Accelerators,
  Spectrometers, Detectors and Associated Equipment} {\bfseries 462} no.~1,
  (2001) 152 -- 155}.
  \url{http://www.sciencedirect.com/science/article/pii/S0168900201000894}.

\bibitem{Agostinelli:2002hh}
{\bfseries GEANT4}, S.~Agostinelli {\em et~al.}, ``{GEANT4: A Simulation
  toolkit}'',
\href{http://dx.doi.org/https://doi.org/10.1016/S0168-9002(03)01368-8}{{\em
  Nucl.Instrum.Meth.} {\bfseries A506} (2003) 250--303}.

\bibitem{BARBERIO1991115}
E.~Barberio {\em et~al.}, ``{Photos — a universal Monte Carlo for QED
  radiative corrections in decays}'',
  \href{http://dx.doi.org/https://doi.org/10.1016/0010-4655(91)90012-A}{{\em
  Computer Physics Communications} {\bfseries 66} no.~1, (1991) 115 -- 128}.
  \url{http://www.sciencedirect.com/science/article/pii/001046559190012A}.

\bibitem{Ward:2002qq}
B.~Ward, S.~Jadach, and Z.~Was, ``{Precision calculation for \HepProcess{
  \Ppositron \Pelectron \to 2\Pfermion}: The KKMC project}'',
  \href{http://dx.doi.org/10.1016/S0920-5632(03)80147-0}{{\em Nucl. Phys. B
  Proc. Suppl.} {\bfseries 116} (2003) 73--77},
  \href{http://arxiv.org/abs/hep-ph/0211132}{{\ttfamily arXiv:hep-ph/0211132}}.

\bibitem{Sj_strand_2008}
T.~Sjöstrand {\em et~al.}, ``{A brief introduction to PYTHIA 8.1}'',
  \href{http://dx.doi.org/10.1016/j.cpc.2008.01.036}{{\em Computer Physics
  Communications} {\bfseries 178} no.~11, (Jun, 2008) 852–867}.
  \url{http://dx.doi.org/10.1016/j.cpc.2008.01.036}.

\bibitem{Kuhr:2018lps}
{\bfseries Belle II Framework Software Group}, T.~Kuhr {\em et~al.}, ``{The
  Belle II Core Software}'',
  \href{http://dx.doi.org/10.1007/s41781-018-0017-9}{{\em Comput. Softw. Big
  Sci.} {\bfseries 3} no.~1, (2019) 1},
  \href{http://arxiv.org/abs/1809.04299}{{\ttfamily arXiv:1809.04299}}.

\bibitem{Bertacchi:2020eez}
{\bfseries Belle II Tracking}, V.~Bertacchi {\em et~al.}, ``{Track Finding at
  Belle II}'',  (3, 2020) , \href{http://arxiv.org/abs/2003.12466}{{\ttfamily
  arXiv:2003.12466}}.

\bibitem{Boyd_1997}
C.~Boyd {\em et~al.}, ``{Precision corrections to dispersive bounds on form
  factors}'', \href{http://dx.doi.org/10.1103/physrevd.56.6895}{{\em Physical
  Review D} {\bfseries 56} no.~11, (Dec, 1997) 6895–6911}.
  \url{http://dx.doi.org/10.1103/PhysRevD.56.6895}.

\bibitem{Glattauer_2016}
{\bfseries Belle}, R.~Glattauer {\em et~al.}, ``{Measurement of the decay
  \BtoDlnu in fully reconstructed events and determination of the
  Cabibbo-Kobayashi-Maskawa matrix element $|V_{cb}|$}'',
  \href{http://dx.doi.org/10.1103/physrevd.93.032006}{{\em Physical Review D}
  {\bfseries 93} no.~3, (Feb, 2016) }.
  \url{http://dx.doi.org/10.1103/PhysRevD.93.032006}.

\bibitem{CAPRINI1998153}
I.~Caprini {\em et~al.}, ``{Dispersive bounds on the shape of \HepProcess{\PB
  \to \PDst \Plepton \Pgnl} form factors}'',
  \href{http://dx.doi.org/https://doi.org/10.1016/S0550-3213(98)00350-2}{{\em
  Nuclear Physics B} {\bfseries 530} no.~1, (1998) 153 -- 181}.

\bibitem{Amhis_2017}
Y.~Amhis {\em et~al.}, ``{Averages of \Pqb-hadron, \Pqc-hadron, and
  \Ptau-lepton properties as of summer 2016}'',
  \href{http://dx.doi.org/10.1140/epjc/s10052-017-5058-4}{{\em The European
  Physical Journal C} {\bfseries 77} no.~12, (Dec, 2017) }.

\bibitem{Leibovich_1998}
A.~Leibovich {\em et~al.}, ``{Semileptonic \PB decays to excited charmed
  mesons}'', \href{http://dx.doi.org/10.1103/physrevd.57.308}{{\em Physical
  Review D} {\bfseries 57} no.~1, (Jan, 1998) 308–330}.
  \url{http://dx.doi.org/10.1103/PhysRevD.57.308}.

\bibitem{Bernlochner_2017}
F.~Bernlochner and Z.~Ligeti, ``{Semileptonic $B(s)$ decays to excited charmed
  mesons with e, $\mu$, $\tau$ and searching for new physics with
  $R(D^{**})$}'', \href{http://dx.doi.org/10.1103/physrevd.95.014022}{{\em
  Physical Review D} {\bfseries 95} no.~1, (Jan, 2017) }.
  \url{http://dx.doi.org/10.1103/PhysRevD.95.014022}.

\bibitem{Goity_1995}
J.~Goity and W.~Roberts, ``{Soft pion emission in semileptonic \PB-meson
  decays}'', \href{http://dx.doi.org/10.1103/physrevd.51.3459}{{\em Physical
  Review D} {\bfseries 51} no.~7, (Apr, 1995) 3459–3477}.
  \url{http://dx.doi.org/10.1103/PhysRevD.51.3459}.

\bibitem{Fox:1978vu}
G.~Fox and S.~Wolfram, ``{Observables for the Analysis of Event Shapes in
  \Ppositron\Pelectron Annihilation and Other Processes}'',
\href{http://dx.doi.org/10.1103/PhysRevLett.41.1581}{{\em Phys. Rev. Lett.}
  {\bfseries 41} (1978) 1581}.

\bibitem{Hershenhorn:454}
A.~Hershenhorn {\em et~al.}, ``{ECL shower shape variables based on Zernike
  moments}'', {\em Internal Note} (Jan, 2017) .

\bibitem{PhysRevD.53.1039}
D.~Asner {\em et~al.}, ``{Search for exclusive charmless hadronic \PB
  decays}'', \href{http://dx.doi.org/10.1103/PhysRevD.53.1039}{{\em Phys. Rev.
  D} {\bfseries 53} (Feb, 1996) 1039--1050}.

\bibitem{Bevan_2014}
A.~J. Bevan {\em et~al.}, ``{The Physics of the B Factories}'',
  \href{http://dx.doi.org/10.1140/epjc/s10052-014-3026-9}{{\em The European
  Physical Journal C} {\bfseries 74} no.~11, (Nov, 2014) }.
  \url{http://dx.doi.org/10.1140/epjc/s10052-014-3026-9}.

\bibitem{Keck:2017gsv}
T.~Keck, ``{FastBDT: A Speed-Optimized Multivariate Classification Algorithm
  for the Belle II Experiment}'',
  \href{http://dx.doi.org/10.1007/s41781-017-0002-8}{{\em Comput. Softw. Big
  Sci.} {\bfseries 1} no.~1, (2017) 2}.
  \url{https://doi.org/10.1007/s41781-017-0002-8}.

\bibitem{Sutcliffe:1470}
W.~Sutcliffe, ``{Performance of Full Event Interpretation and a calibration
  with $B \rightarrow X \ell \nu$ decays in early phase III data}'', {\em
  Internal Note} (Jul, 2019) .

\bibitem{GATZ19951185}
D.~Gatz and L.~Smith, ``{The standard error of a weighted mean concentration.
  Bootstrapping vs other methods}'',
  \href{http://dx.doi.org/https://doi.org/10.1016/1352-2310(94)00210-C}{{\em
  Atmospheric Environment} {\bfseries 29} no.~11, (1995) 1185 -- 1193}.
  \url{http://www.sciencedirect.com/science/article/pii/135223109400210C}.

\bibitem{PhysRevD.98.030001}
{\bfseries Particle Data Group}, M.~Tanabashi {\em et~al.}, ``{Review of
  Particle Physics}'', \href{http://dx.doi.org/10.1103/PhysRevD.98.030001}{{\em
  Phys. Rev. D} {\bfseries 98} (Aug, 2018) 030001}.

\bibitem{Aubert:2007yaa}
{\bfseries BaBar}, B.~Aubert {\em et~al.}, ``{Measurement of moments of the
  hadronic-mass and energy spectrum in inclusive semileptonic \BtoXclv}'', in
  {\em {2007 Europhysics Conference on High Energy Physics}}.
\newblock 7, 2007.
\newblock \href{http://arxiv.org/abs/0707.2670}{{\ttfamily arXiv:0707.2670}}.

\bibitem{Schwanda_2007}
{\bfseries Belle}, C.~Schwanda {\em et~al.}, ``{Moments of the hadronic
  invariant mass spectrum in \BtoXclv decays at Belle}'',
  \href{http://dx.doi.org/10.1103/physrevd.75.032005}{{\em Physical Review D}
  {\bfseries 75} no.~3, (Feb, 2007) }.
  \url{http://dx.doi.org/10.1103/PhysRevD.75.032005}.

\end{thebibliography}\endgroup
\bibliographystyle{utphys}
\clearpage
\appendix
\onecolumngrid

%

\section{Numerical Results and Breakdown of Statistical and Systematic Uncertainties}
\label{sec:numerical_values}
\tiny
\begin{table}[tb]
    \centering
    \setlength{\tabcolsep}{0.5em}
    \caption{Summary of statistical and systematic uncertainties for the measurement of \mxmoment{}. All values are given in $\mathrm{\giga\eVperc\squared}$ if not stated otherwise. The calculation of the uncertainties is described in
    \cref{subsec:uncertainties}.}
    
    \vspace{0.2cm}
    
    \begin{tabular}{lrrrrrr}
    \toprule
     \plepsigbrestframe Cut in $\mathrm{\giga\eVperc}$ &     0.8 &     0.9 &     1.0 &     1.1 &     1.2 &     1.3 \\
    \colrule
    \mxmoment{} in $\mathrm{\giga\eVperc\squared}$         &  2.1144 &  2.1093 &  2.0996 &  2.0899 &  2.0795 &  2.0729 \\
    \colrule
    Stat. error (data)         &  0.0035 &  0.0036 &  0.0038 &  0.0039 &  0.0042 &  0.0045 \\
    Stat. error (signal prob.) &  0.0068 &  0.0021 &  0.0013 &  0.0009 &  0.0000 &  0.0003 \\
    \colrule
    Stat. error (total)        &  0.0076 &  0.0042 &  0.0040 &  0.0040 &  0.0042 &  0.0045 \\
    \colrule
    Calib. function error      &  0.0107 &  0.0102 &  0.0099 &  0.0096 &  0.0093 &  0.0090 \\
    FEI eff..                   &  0.0059 &  0.0035 &  0.0020 &  0.0009 &  0.0000 &  0.0004 \\
    PID eff.                   &  0.0086 &  0.0042 &  0.0032 &  0.0022 &  0.0013 &  0.0011 \\
    \BtoXulv BF                &  0.0042 &  0.0041 &  0.0040 &  0.0041 &  0.0042 &  0.0044 \\
    Bias corr. (stat)          &  0.0025 &  0.0025 &  0.0025 &  0.0025 &  0.0026 &  0.0027 \\
    Bias corr. (model)         &  0.0421 &  0.0384 &  0.0345 &  0.0301 &  0.0265 &  0.0237 \\
    \colrule
    Sys. error (total)         &  0.0449 &  0.0404 &  0.0364 &  0.0320 &  0.0285 &  0.0260 \\
    \colrule
    Total error                &  0.0456 &  0.0406 &  0.0366 &  0.0323 &  0.0289 &  0.0264 \\
    \toprule
     \plepsigbrestframe Cut in $\mathrm{\giga\eVperc}$   &     1.4 &     1.5 &     1.6 &     1.7 &     1.8 &     1.9 \\
    \colrule
    \mxmoment{} in $\mathrm{\giga\eVperc\squared}$         &  2.0641 &  2.0592 &  2.0502 &  2.0366 &  2.0386 &  2.0429 \\
    \colrule
    Stat. error (data)         &  0.0050 &  0.0057 &  0.0066 &  0.0082 &  0.0103 &  0.0132 \\
    Stat. error (signal prob.) &  0.0008 &  0.0007 &  0.0009 &  0.0018 &  0.0028 &  0.0042 \\
    \colrule
    Stat. error (total)        &  0.0051 &  0.0057 &  0.0067 &  0.0084 &  0.0107 &  0.0139 \\
    \colrule
    Calib. function error      &  0.0088 &  0.0086 &  0.0083 &  0.0074 &  0.0077 &  0.0076 \\
    FEI eff..                   &  0.0008 &  0.0012 &  0.0015 &  0.0019 &  0.0026 &  0.0037 \\
    PID eff.                   &  0.0009 &  0.0008 &  0.0009 &  0.0011 &  0.0014 &  0.0019 \\
    \BtoXulv BF                &  0.0048 &  0.0054 &  0.0067 &  0.0083 &  0.0101 &  0.0142 \\
    Bias corr. (stat)          &  0.0029 &  0.0033 &  0.0037 &  0.0045 &  0.0057 &  0.0075 \\
    Bias corr. (model)         &  0.0200 &  0.0168 &  0.0139 &  0.0109 &  0.0074 &  0.0042 \\
    \colrule
    Sys. error (total)         &  0.0226 &  0.0200 &  0.0180 &  0.0164 &  0.0161 &  0.0187 \\
    \colrule
    Total error                &  0.0232 &  0.0208 &  0.0192 &  0.0184 &  0.0193 &  0.0233 \\
    \botrule
    \end{tabular}
    \label{tab:mx_uncertainties}
\end{table}

\begin{table}
    \centering
    \setlength{\tabcolsep}{0.5em}
    \caption{Summary of statistical and systematic uncertainties for the measurement of \mxmoment{2}. All values are given in $(\mathrm{\giga\eVperc\squared})^2$ if not stated otherwise. The calculation of the uncertainties is described in
    \cref{subsec:uncertainties}.}
    
    \vspace{0.2cm}
    
    \begin{tabular}{lrrrrrr}
    \toprule
     \plepsigbrestframe Cut in $\mathrm{\giga\eVperc}$   &     0.8 &     0.9 &     1.0 &     1.1 &     1.2 &     1.3 \\
    \colrule
    \mxmoment{2} in $(\mathrm{\giga\eVperc\squared})^2$         &  4.5743 &  4.5459 &  4.4902 &  4.4365 &  4.3790 &  4.3458 \\
    \colrule
    Stat. error (data)         &  0.0146 &  0.0151 &  0.0157 &  0.0165 &  0.0175 &  0.0189 \\
    Stat. error (signal prob.) &  0.0405 &  0.0140 &  0.0092 &  0.0071 &  0.0017 &  0.0003 \\
    \colrule
    Stat. error (total)        &  0.0431 &  0.0206 &  0.0182 &  0.0180 &  0.0176 &  0.0189 \\
    \colrule
    Calib. function error      &  0.0473 &  0.0447 &  0.0427 &  0.0410 &  0.0393 &  0.0380 \\
    FEI eff..                   &  0.0340 &  0.0201 &  0.0118 &  0.0060 &  0.0014 &  0.0005 \\
    PID eff.                   &  0.0476 &  0.0210 &  0.0164 &  0.0109 &  0.0060 &  0.0046 \\
    \BtoXulv BF                &  0.0168 &  0.0157 &  0.0151 &  0.0150 &  0.0153 &  0.0160 \\
    Bias corr. (stat)          &  0.0115 &  0.0112 &  0.0110 &  0.0110 &  0.0112 &  0.0116 \\
    Bias corr. (model)         &  0.2099 &  0.1902 &  0.1687 &  0.1446 &  0.1254 &  0.1106 \\
    \colrule
    Sys. error (total)         &  0.2239 &  0.1985 &  0.1762 &  0.1519 &  0.1329 &  0.1187 \\
    \colrule
    Total error                &  0.2280 &  0.1996 &  0.1771 &  0.1530 &  0.1340 &  0.1202 \\
    \toprule
    \plepsigbrestframe Cut in $\mathrm{\giga\eVperc}$   &     1.4 &     1.5 &     1.6 &     1.7 &     1.8 &     1.9 \\
    \colrule
    \mxmoment{2} in $(\mathrm{\giga\eVperc\squared})^2$        &  4.2980 &  4.2691 &  4.2209 &  4.1483 &  4.1493 &  4.1547 \\
    \colrule
    Stat. error (data)         &  0.0208 &  0.0235 &  0.0274 &  0.0337 &  0.0426 &  0.0553 \\
    Stat. error (signal prob.) &  0.0011 &  0.0017 &  0.0026 &  0.0054 &  0.0088 &  0.0137 \\
    \colrule
    Stat. error (total)        &  0.0208 &  0.0236 &  0.0275 &  0.0341 &  0.0435 &  0.0570 \\
    \colrule
    Calib. function error      &  0.0366 &  0.0355 &  0.0339 &  0.0296 &  0.0310 &  0.0303 \\
    FEI eff..                   &  0.0020 &  0.0038 &  0.0050 &  0.0065 &  0.0092 &  0.0134 \\
    PID eff.                   &  0.0037 &  0.0032 &  0.0035 &  0.0041 &  0.0051 &  0.0070 \\
    \BtoXulv BF                &  0.0171 &  0.0200 &  0.0228 &  0.0283 &  0.0358 &  0.0503 \\
    Bias corr. (stat)          &  0.0123 &  0.0135 &  0.0154 &  0.0184 &  0.0230 &  0.0303 \\
    Bias corr. (model)         &  0.0920 &  0.0764 &  0.0621 &  0.0483 &  0.0328 &  0.0185 \\
    \colrule
    Sys. error (total)         &  0.1013 &  0.0878 &  0.0761 &  0.0664 &  0.0629 &  0.0703 \\
    \colrule
    Total error                &  0.1034 &  0.0909 &  0.0810 &  0.0746 &  0.0765 &  0.0905 \\
    \botrule
    \end{tabular}
    \label{tab:mx2_uncertainties}
\end{table}

\begin{table}
    \centering
    \setlength{\tabcolsep}{0.5em}
    \caption{Summary of statistical and systematic uncertainties for the measurement of \mxmoment{3}. All values are given in $(\mathrm{\giga\eVperc\squared})^3$ if not stated otherwise. The calculation of the uncertainties is described in
    \cref{subsec:uncertainties}.}
    
    \vspace{0.2cm}
    
    \begin{tabular}{lrrrrrr}
    \toprule
     \plepsigbrestframe Cut in $\mathrm{\giga\eVperc}$ &      0.8 &     0.9 &     1.0 &     1.1 &     1.2 &     1.3 \\
    \colrule
    \mxmoment{3} in $(\mathrm{\giga\eVperc\squared})^3$         &  10.2132 &  10.0919 &  9.8513 &  9.6251 &  9.3849 &  9.2553 \\
    \colrule
    Stat. error (data)         &   0.0475 &   0.0492 &  0.0509 &  0.0534 &  0.0564 &  0.0608 \\
    Stat. error (signal prob.) &   0.1830 &   0.0645 &  0.0431 &  0.0344 &  0.0108 &  0.0054 \\
    \colrule
    Stat. error (total)        &   0.1891 &   0.0811 &  0.0667 &  0.0635 &  0.0574 &  0.0610 \\
    \colrule
    Calib. function error      &   0.1668 &   0.1556 &  0.1463 &  0.1383 &  0.1302 &  0.1250 \\
    FEI eff..                   &   0.1493 &   0.0875 &  0.0517 &  0.0273 &  0.0088 &  0.0019 \\
    PID eff.                   &   0.2065 &   0.0788 &  0.0660 &  0.0422 &  0.0210 &  0.0153 \\
    \BtoXulv BF                &   0.0535 &   0.0485 &  0.0448 &  0.0435 &  0.0435 &  0.0452 \\
    Bias corr. (stat)          &   0.0429 &   0.0407 &  0.0391 &  0.0382 &  0.0377 &  0.0384 \\
    Bias corr. (model)         &   0.8077 &   0.7266 &  0.6339 &  0.5331 &  0.4533 &  0.3929 \\
    \colrule
    Sys. error (total)         &   0.8659 &   0.7550 &  0.6586 &  0.5560 &  0.4756 &  0.4168 \\
    \colrule
    Total error                &   0.8863 &   0.7594 &  0.6620 &  0.5596 &  0.4791 &  0.4213 \\
    \toprule
    \plepsigbrestframe Cut in $\mathrm{\giga\eVperc}$  &     1.4 &     1.5 &     1.6 &     1.7 &     1.8 &     1.9 \\
    \colrule
    \mxmoment{3} in $(\mathrm{\giga\eVperc\squared})^3$        &  9.0639 &  8.9409 &  8.7514 &  8.4779 &  8.4616 &  8.4534 \\
    \colrule
    Stat. error (data)         &  0.0664 &  0.0749 &  0.0867 &  0.1056 &  0.1339 &  0.1746 \\
    Stat. error (signal prob.) &  0.0016 &  0.0030 &  0.0055 &  0.0116 &  0.0210 &  0.0347 \\
    \colrule
    Stat. error (total)        &  0.0664 &  0.0750 &  0.0869 &  0.1063 &  0.1355 &  0.1780 \\
    \colrule
    Calib. function error      &  0.1186 &  0.1140 &  0.1073 &  0.0919 &  0.0961 &  0.0932 \\
    FEI eff..                   &  0.0036 &  0.0093 &  0.0131 &  0.0175 &  0.0250 &  0.0367 \\
    PID eff.                   &  0.0118 &  0.0093 &  0.0102 &  0.0118 &  0.0143 &  0.0195 \\
    \BtoXulv BF                &  0.0476 &  0.0565 &  0.0617 &  0.0761 &  0.0978 &  0.1373 \\
    Bias corr. (stat)          &  0.0399 &  0.0434 &  0.0487 &  0.0572 &  0.0716 &  0.0940 \\
    Bias corr. (model)         &  0.3208 &  0.2624 &  0.2100 &  0.1604 &  0.1084 &  0.0607 \\
    \colrule
    Sys. error (total)         &  0.3478 &  0.2951 &  0.2492 &  0.2090 &  0.1910 &  0.2044 \\
    \colrule
    Total error                &  0.3541 &  0.3045 &  0.2639 &  0.2345 &  0.2342 &  0.2711 \\
    \botrule
    \end{tabular}
    \label{tab:mx3_uncertainties}
\end{table}

\begin{table}
    \centering
    \setlength{\tabcolsep}{0.5em}
    \caption{Summary of statistical and systematic uncertainties for the measurement of \mxmoment{4}. All values are given in $(\mathrm{\giga\eVperc\squared})^4$ if not stated otherwise. The calculation of the uncertainties is described in
    \cref{subsec:uncertainties}.}
    
    \vspace{0.2cm}
    
    \begin{tabular}{lrrrrrr}
    \toprule
    \plepsigbrestframe Cut in $\mathrm{\giga\eVperc}$ &      0.8 &      0.9 &      1.0 &      1.1 &      1.2 &      1.3 \\
    \colrule
    \mxmoment{4} in $(\mathrm{\giga\eVperc\squared})^4$         &  23.7733 &  23.2997 &  22.3539 &  21.4874 &  20.5818 &  20.1196 \\
    \colrule
    Stat. error (data)         &   0.1420 &   0.1471 &   0.1516 &   0.1584 &   0.1662 &   0.1788 \\
    Stat. error (signal prob.) &   0.7534 &   0.2620 &   0.1742 &   0.1397 &   0.0472 &   0.0276 \\
    \colrule
    Stat. error (total)        &   0.7667 &   0.3005 &   0.2309 &   0.2113 &   0.1728 &   0.1809 \\
    \colrule
    Calib. function error      &   0.5569 &   0.5112 &   0.4709 &   0.4359 &   0.4010 &   0.3808 \\
    FEI eff..                   &   0.5999 &   0.3444 &   0.2012 &   0.1073 &   0.0386 &   0.0150 \\
    PID eff.                   &   0.8303 &   0.2671 &   0.2454 &   0.1511 &   0.0684 &   0.0474 \\
    \BtoXulv BF                &   0.1629 &   0.1425 &   0.1257 &   0.1182 &   0.1146 &   0.1178 \\
    Bias corr. (stat)          &   0.1524 &   0.1406 &   0.1308 &   0.1238 &   0.1183 &   0.1178 \\
    Bias corr. (model)         &   2.8491 &   2.5472 &   2.1796 &   1.7933 &   1.4891 &   1.2646 \\
    \colrule
    Sys. error (total)         &   3.0865 &   2.6419 &   2.2597 &   1.8626 &   1.5529 &   1.3321 \\
    \colrule
    Total error                &   3.1803 &   2.6590 &   2.2714 &   1.8746 &   1.5624 &   1.3444 \\
    \toprule
    \plepsigbrestframe Cut in $\mathrm{\giga\eVperc}$  &      1.4 &      1.5 &      1.6 &      1.7 &      1.8 &      1.9 \\
    \colrule
    \mxmoment{4} in $(\mathrm{\giga\eVperc\squared})^4$         &  19.4346 &  18.9820 &  18.3187 &  17.4161 &  17.3199 &  17.2427 \\
    \colrule
    Stat. error (data)         &   0.1935 &   0.2177 &   0.2487 &   0.2993 &   0.3791 &   0.4942 \\
    Stat. error (signal prob.) &   0.0178 &   0.0026 &   0.0093 &   0.0209 &   0.0449 &   0.0801 \\
    \colrule
    Stat. error (total)        &   0.1943 &   0.2177 &   0.2488 &   0.3000 &   0.3817 &   0.5006 \\
    \colrule
    Calib. function error      &   0.3546 &   0.3360 &   0.3110 &   0.2587 &   0.2695 &   0.2597 \\
    FEI eff..                   &   0.0032 &   0.0205 &   0.0309 &   0.0423 &   0.0619 &   0.0915 \\
    PID eff.                   &   0.0343 &   0.0248 &   0.0268 &   0.0306 &   0.0367 &   0.0492 \\
    \BtoXulv BF                &   0.1218 &   0.1459 &   0.1538 &   0.1884 &   0.2431 &   0.3400 \\
    Bias corr. (stat)          &   0.1195 &   0.1277 &   0.1407 &   0.1615 &   0.2013 &   0.2633 \\
    Bias corr. (model)         &   1.0099 &   0.8108 &   0.6371 &   0.4755 &   0.3194 &   0.1774 \\
    \colrule
    Sys. error (total)         &   1.0844 &   0.8994 &   0.7401 &   0.5978 &   0.5286 &   0.5428 \\
    \colrule
    Total error                &   1.1016 &   0.9254 &   0.7808 &   0.6689 &   0.6520 &   0.7384 \\
    \botrule
    \end{tabular}
    \label{tab:mx4_uncertainties}
\end{table}

\begin{table}
    \centering
    \setlength{\tabcolsep}{0.5em}
    \caption{Summary of statistical and systematic uncertainties for the measurement of \mxmoment{5}. All values are given in $(\mathrm{\giga\eVperc\squared})^5$ if not stated otherwise. The calculation of the uncertainties is described in
    \cref{subsec:uncertainties}.}
    
    \vspace{0.2cm}
    
    \begin{tabular}{lrrrrrr}
    \toprule
    \plepsigbrestframe Cut in $\mathrm{\giga\eVperc}$ &      0.8 &      0.9 &      1.0 &      1.1 &      1.2 &      1.3 \\
    \colrule
   \mxmoment{5} in $(\mathrm{\giga\eVperc\squared})^5$         &  58.2926 &  56.5135 &  52.9344 &  49.7378 &  46.4718 &  44.8842 \\
   \colrule
    Stat. error (data)         &   0.4142 &   0.4295 &   0.4394 &   0.4566 &   0.4749 &   0.5093 \\
    Stat. error (signal prob.) &   3.0074 &   1.0155 &   0.6627 &   0.5267 &   0.1790 &   0.1105 \\
    \colrule
    Stat. error (total)        &   3.0357 &   1.1026 &   0.7951 &   0.6971 &   0.5075 &   0.5211 \\
    \colrule
    Calib. function error      &   1.8603 &   1.6787 &   1.5072 &   1.3584 &   1.2127 &   1.1360 \\
    FEI eff..                   &   2.3394 &   1.3060 &   0.7459 &   0.3943 &   0.1464 &   0.0681 \\
    PID eff.                   &   3.2669 &   0.8661 &   0.8898 &   0.5269 &   0.2171 &   0.1429 \\
    \BtoXulv BF                &   0.4995 &   0.4215 &   0.3507 &   0.3165 &   0.2955 &   0.2991 \\
    Bias corr. (stat)          &   0.5448 &   0.4884 &   0.4375 &   0.3987 &   0.3652 &   0.3539 \\
    Bias corr. (model)         &   9.7284 &   8.6597 &   7.2503 &   5.8219 &   4.7004 &   3.9025 \\
    \colrule
    Sys. error (total)         &  10.7142 &   8.9822 &   7.5167 &   6.0359 &   4.8840 &   4.0939 \\
    \colrule
    Total error                &  11.1360 &   9.0496 &   7.5586 &   6.0760 &   4.9103 &   4.1269 \\
    \toprule
    \plepsigbrestframe Cut in $\mathrm{\giga\eVperc}$  &      1.4 &      1.5 &      1.6 &      1.7 &      1.8 &      1.9 \\
    \colrule
    \mxmoment{5} in $(\mathrm{\giga\eVperc\squared})^5$        &  42.5549 &  41.0086 &  38.8121 &  36.0142 &  35.6291 &  35.2999 \\
    \colrule
    Stat. error (data)         &   0.5452 &   0.6100 &   0.6834 &   0.8081 &   1.0206 &   1.3258 \\
    Stat. error (signal prob.) &   0.0828 &   0.0071 &   0.0106 &   0.0296 &   0.0888 &   0.1766 \\
    \colrule
    Stat. error (total)        &   0.5514 &   0.6101 &   0.6835 &   0.8086 &   1.0245 &   1.3375 \\
    \colrule
    Calib. function error      &   1.0333 &   0.9615 &   0.8690 &   0.6969 &   0.7215 &   0.6895 \\
    FEI eff..                   &   0.0099 &   0.0414 &   0.0691 &   0.0975 &   0.1463 &   0.2178 \\
    PID eff.                   &   0.0972 &   0.0637 &   0.0673 &   0.0762 &   0.0895 &   0.1183 \\
    \BtoXulv BF                &   0.3015 &   0.3649 &   0.3684 &   0.4490 &   0.5789 &   0.8052 \\
    Bias corr. (stat)          &   0.3473 &   0.3629 &   0.3897 &   0.4350 &   0.5390 &   0.7018 \\
    Bias corr. (model)         &   3.0350 &   2.3830 &   1.8298 &   1.3269 &   0.8852 &   0.4869 \\
    \colrule
    Sys. error (total)         &   3.2404 &   2.6218 &   2.0976 &   1.6286 &   1.3997 &   1.3837 \\
    \colrule
    Total error                &   3.2870 &   2.6919 &   2.2062 &   1.8183 &   1.7346 &   1.9245 \\
    \botrule
    \end{tabular}
    \label{tab:mx5_uncertainties}
\end{table}

\begin{table}
    \centering
    \setlength{\tabcolsep}{0.5em}
    \caption{Summary of statistical and systematic uncertainties for the measurement of \mxmoment{6}. All values are given in $(\mathrm{\giga\eVperc\squared})^6$ if not stated otherwise. The calculation of the uncertainties is described in
    \cref{subsec:uncertainties}.}
    
    \vspace{0.2cm}
    
    \begin{tabular}{lrrrrrr}
    \toprule
    \plepsigbrestframe Cut in $\mathrm{\giga\eVperc}$ &      0.8 &      0.9 &      1.0 &      1.1 &      1.2 &      1.3 \\
    \colrule
   \mxmoment{6} in $(\mathrm{\giga\eVperc\squared})^6$         &  151.8801 &  145.3258 &  131.9459 &  120.3054 &  108.7374 &  103.3617 \\
   \colrule
    Stat. error (data)         &    1.2115 &    1.2581 &    1.2752 &    1.3148 &    1.3525 &    1.4462 \\
    Stat. error (signal prob.) &   11.9493 &    3.8818 &    2.4632 &    1.9239 &    0.6386 &    0.3983 \\
    \colrule
    Stat. error (total)        &   12.0106 &    4.0806 &    2.7737 &    2.3302 &    1.4956 &    1.5001 \\
    \colrule
    Calib. function error      &    6.3730 &    5.6553 &    4.9278 &    4.2983 &    3.6992 &    3.4080 \\
    FEI eff..                   &    9.0921 &    4.9122 &    2.7203 &    1.4068 &    0.5192 &    0.2599 \\
    PID eff.                   &   12.8615 &    2.7396 &    3.2193 &    1.8326 &    0.6869 &    0.4299 \\
    \BtoXulv BF                &    1.5766 &    1.2883 &    1.0012 &    0.8586 &    0.7635 &    0.7571 \\
    Bias corr. (stat)          &    1.9994 &    1.7482 &    1.5011 &    1.3090 &    1.1407 &    1.0690 \\
    Bias corr. (model)         &   32.9241 &   29.2623 &   23.9298 &   18.7205 &   14.6357 &   11.8590 \\
    \colrule
    Sys. error (total)         &   37.1373 &   30.4075 &   24.8584 &   19.4093 &   15.1827 &   12.4185 \\
    \colrule
    Total error                &   39.0312 &   30.6801 &   25.0127 &   19.5487 &   15.2562 &   12.5088 \\
    \toprule
    \plepsigbrestframe Cut in $\mathrm{\giga\eVperc}$  &      1.4 &      1.5 &      1.6 &      1.7 &      1.8 &      1.9 \\
    \colrule
    \mxmoment{6} in $(\mathrm{\giga\eVperc\squared})^6$         &  95.6289 &  90.5528 &  83.4604 &  75.0624 &  73.7412 &  72.5957 \\
    \colrule
    Stat. error (data)         &   1.5273 &   1.6988 &   1.8468 &   2.1309 &   2.6788 &   3.4604 \\
    Stat. error (signal prob.) &   0.3134 &   0.0532 &   0.0050 &   0.0173 &   0.1625 &   0.3796 \\
    \colrule
    Stat. error (total)        &   1.5591 &   1.6997 &   1.8468 &   2.1310 &   2.6837 &   3.4811 \\
    \colrule
    Calib. function error      &   3.0150 &   2.7442 &   2.3999 &   1.8381 &   1.8872 &   1.7875 \\
    FEI eff..                   &   0.0746 &   0.0764 &   0.1491 &   0.2192 &   0.3380 &   0.5067 \\
    PID eff.                   &   0.2739 &   0.1611 &   0.1657 &   0.1856 &   0.2133 &   0.2771 \\
    \BtoXulv BF                &   0.7394 &   0.9072 &   0.8658 &   1.0502 &   1.3508 &   1.8659 \\
    Bias corr. (stat)          &   1.0066 &   1.0232 &   1.0616 &   1.1438 &   1.4076 &   1.8225 \\
    Bias corr. (model)         &   8.9427 &   6.8406 &   5.1028 &   3.5697 &   2.3624 &   1.2839 \\
    \colrule
    Sys. error (total)         &   9.5238 &   7.4984 &   5.8073 &   4.3146 &   3.6205 &   3.4612 \\
    \colrule
    Total error                &   9.6506 &   7.6887 &   6.0939 &   4.8121 &   4.5067 &   4.9090 \\
    \botrule
    \end{tabular}
    \label{tab:mx6_uncertainties}
\end{table}

\end{document}